\documentclass[letterpaper,prd,preprint,floatfix,showpacs]{revtex4-1}
\usepackage[pdftex]{color}
\usepackage[dvipsnames]{xcolor}
\usepackage{latexsym,amssymb,amsmath,gensymb}
\usepackage{graphicx,dcolumn,bm}
\usepackage{kec,bigints}
\usepackage{mathtools}
\usepackage{float}
\usepackage{caption}
\begin{document}

\title{Path integrals for awkward actions}

\author{David Amdahl} 
\email{damdahl@unm.edu}
\author{Kevin Cahill}
\email{cahill@unm.edu}
\affiliation{Department of Physics \& Astronomy,
University of New Mexico, Albuquerque, New Mexico 
87131, USA}
\affiliation{School of Computational Sciences,
Korea Institute for Advanced Study, Seoul 130-722, Korea}

\date{\today}

\begin {abstract}
Time derivatives of scalar fields
occur quadratically in textbook actions.
A simple Legendre transformation 
turns the lagrangian into a 
hamiltonian that is quadratic in the momenta.
The path integral over the momenta is gaussian.
Mean values of operators are 
euclidian path integrals of their classical counterparts
with positive weight functions.
Monte Carlo simulations can estimate 
such mean values.
\par
This familiar framework falls apart
when the time derivatives do not occur
quadratically.  
The Legendre transformation 
becomes difficult or so intractable
that one can't find the hamiltonian.  
Even if one finds the hamiltonian, 
it usually is so complicated
that one can't path-integrate 
over the momenta and get 
a euclidian path integral with
a positive weight function.
Monte Carlo simulations don't work
when the weight function
assumes negative or complex values.
\par
This paper solves both problems.
It shows how to make path integrals 
without knowing the hamiltonian. 
It also shows how to estimate
complex path integrals 
by combining the Monte Carlo method
with parallel numerical integration and
a lookup table. 	
This ``Atlantic City" method lets one
estimate the energy densities
of theories that, unlike those 
with quadratic time derivatives,
may have finite energy densities.
It may lead to a theory of dark energy.
\par 
The approximation of multiple
integrals over weight functions
that assume negative or complex values
is the long-standing sign problem.
The Atlantic City method
solves it for problems
in which numerical integration
leads to a positive weight function.
\end {abstract}

\maketitle

\section {Introduction
\label {Introduction} }
Despite the success of renormalization,
infinities remain a major problem
in quantum field theory. 
This problem
grows more important
as cosmological observations
continue to support the existence of 
dark energy~\cite{PlanckCosmologicalNA},
which may be the energy density 
of empty space.
We need to be able to compute 
finite energy densities. 
This paper advances
theories of scalar fields 
a step closer to that goal.
\par
The ground-state energy
of a theory
is the low-temperature limit 
of the logarithmic derivative 
of the partition function \( Z(\beta) \)
with respect to the inverse temperature \( \beta \)\@.
If the action density \( L \) is quadratic
in the time derivatives 
\( \dot \phi = \dot \phi_1, \dots, \dot \phi_n\) 
of the fields, then
a linear Legendre transformation
gives a hamiltonian \( H \) that is
quadratic in the momenta 
\( \pi = \pi_1, \dots, \pi_n \)\@.
One can use the hamiltonian
to write the partition function
as a euclidian path integral
in which the momentum integrals
are gaussian.
Integrating over the momenta,
one gets 
the partition function as 
a path integral of a probability
distribution in the fields.
One then can use Monte Carlo methods
to estimate the partition function
and the mean values of various
observables.
\par
This simple framework falls apart
when the time derivatives do not occur
quadratically.  
This collapse is unfortunate
because theories of scalar fields 
that are quadratic
in the time derivatives of the fields
have infinite energy densities.
\par
An {\bf{awkward}} action is one 
that is not quadratic
in the time derivatives but that is simple
enough for one to find its hamiltonian.
One typically can't integrate over the momentum \( \pi \),
and the partition function is a double path integral
with a complex weight function~\cite{Weinberg1995IX}
\begin{align} 
Z(\beta) ={}& 
\int \exp \lt\{ \int_0^\beta \int \left[ i \dot \phi \pi 
- H(\phi, \pi) \right] dt \, d^3x \rt\} D\phi D\pi .
\label {the path integral for the partition function}
\end{align}
Standard Monte Carlo methods fail
when the weight function assumes negative
or complex values. 

\par
A {\bf{very awkward}} action is one
in which the time derivatives of the fields
are related to their momenta, 
the fields, and their spatial derivatives
by equations that are not even quartic
and so have no algebraic solutions.
Very awkward actions usually
have no known hamiltonians.
To study the ground states
of this wide class of theories,
we show in
Sec.~\ref {Path integrals for very awkward actions}
how to write the partition function
of such a theory as a path integral
without knowledge of the hamiltonian.
Our formula~\cite{Cahill1501} 
is a double path integral
over the fields \( \phi \) and over
auxilliary time derivatives \( \dot \psi \)
\begin{equation}
   \begin{split}
Z(\beta) = {}&
\! \int \exp \bigg\{ \int_0^\beta \! \int 
\Big[ ( i \dot \phi_\ell
- \dot \psi_\ell ) \frac{ \p  L(\phi, \dot \psi) }
{ \p \dot \psi_\ell }
+ L(\phi, \dot \psi) 
\Big] dt d^3x \bigg\}  \lt| \det \! 
\lt( \frac{ \p^2  L(\phi, \dot \psi) }
{ \p \dot \psi_k \p \dot \psi_\ell }  \rt) \rt|
\, D\phi D\dot\psi 
\label {first proposed path integral}
   \end{split}
\end{equation}
in which the \( n \times n \) determinant 
is over the indices \( k = 1, \dots, n \) 
and \( \ell = 1, \dots, n \)
of the fields.
We give four examples
of this formula, in one of which
we incidentally show that 
the classical energy
of the Nambu-Got{\={o}} string
is identically zero.
The path integral (\ref {first proposed path integral}), 
like the one 
(\ref {the path integral for the partition function})
for awkward actions,
has a complex weight function
that is not a probability distribution.
Again the usual Monte Carlo methods
do not work.  
Both path integrals are examples
of what has been called the sign problem.
\par
To estimate such complex path integrals,
we introduce in 
Sec.\,\ref{The Atlantic City method}
a way that combines the 
Monte Carlo method with parallel numerical integration
and lookup tables.
In theories with awkward actions,
we numerically integrate over the momenta
\( \pi \) in the double path integral 
(\ref {the path integral for the partition function})\@.
In theories with very awkward actions,
we numerically integrate over the auxilliary
time derivatives \( \dot \psi \) in the double path integral 
(\ref {first proposed path integral})\@.
In both cases, we store the values of the integrals
in lookup tables and then 
use the lookup tables
to guide standard Monte Carlo estimates.
We call this the Atlantic City way.
It is well suited to
parallel computation and may solve
some versions of the sign 
problem.
We demonstrate and test the method
by applying it to a
quantum-mechanical version of the scalar Born-Infeld 
model~\cite{Born:1934gh, *Born:1934dia, *Born:1935ap}
considered as a theory with an awkward action
in
Sec.\,\ref {Application of the Atlantic City method to the Born-Infeld oscillator} and as a theory with a very awkward action in
Sec.\,\ref{The Atlantic City model applied to a very awkward action}\@.
In Sec.\,\ref{Transition to field theory},
we extend the Atlantic City way 
to field theory and use it to estimate
the known Green's functions of the scalar
free field theory.  
The paper ends with a summary (Sec.\,\ref{Summary})
and an appendix.
\par
The paper does not discuss theories of fields with non-zero spin or
higher
derivatives~\cite{BenderMannheim2007, *
BenderMannheim2008} 
or those in which some 
fields have no time derivatives
\cite{Dirac1950, *Dirac1958, *Dirac1964}\@.

\section {Review of Legendre transformations and path integrals
\label {Review of Legendre transformations and path integrals} }

The lagrangian of a theory tells us about symmetries
and equations of motion, but one needs
a hamiltonian to determine 
the time evolution of states and their energies. 
To find the hamiltonian of a theory
of scalar fields 
\( \phi = \{\phi_1,\dots, \phi_n\} \),
one defines the conjugate momenta 
\( \pi = \{\pi_1,\dots, \pi_n\} \) as
the derivatives of the action density
\begin{equation}
\pi_j = \frac{\p  L }{\p \dot \phi_j} ,
\label {def of pi}
\end{equation}
and inverts these equations so as to
write the time derivatives 
\( \dot \phi_j = \dot \phi_j (\phi, \pi) \)
of the fields in terms of 
the fields \( \phi_\ell \) 
(and possibly their spatial derivatives)
and their momenta \( \pi_\ell \)\@.
The hamiltonian density then is
\begin{equation}
H = \sum_{j=1}^n \pi_j \dot \phi_j(\phi,\pi) 
- L (\phi,\dot \phi(\phi,\pi) ) .
\label {energy density}
\end{equation}
When the action is quadratic 
in the time derivatives,
Legendre's equations (\ref {def of pi})
are linear.  
\par
Once one has a hamiltonian,
one inserts complete sets
of eigenstates of the fields \( \phi_j \) and 
their conjugate momenta \( \pi_j \)
into the Boltzmann operator 
\( \exp( - \beta H ) = ( \exp( - \beta H/n ) )^n \)
and writes the partition function as
the complex path integral~\cite{Weinberg1995IX}
\begin{equation}
Z(\beta) ={} \Tr \, e^{-\beta H}
= \int \la \phi | e^{- \beta H} | \phi \ra D\phi = 
\int \exp \lt\{ \int_0^\beta \int \left[ i \dot \phi_j \pi_j 
- H(\phi, \pi) \right] dt \, d^3x \rt\} D\phi D\pi .
\label {leads to the partition function}
\end{equation}
If one can integrate over the momenta,
then one gets Feynman's formula~\cite{Weinberg1995IX, CahillXVI}
\begin{equation}
Z(\beta) ={} \int \exp 
\lt[ \int_0^\beta \int 
{} - L_e(\phi, \dot \phi) \, dt \, d^3x \rt] D\phi 
\label {one gets the classic formula}
\end{equation}
in which \( L_e \) is the euclidian
action density, and \( D\phi \) is suitably
redefined.
In textbook theories, \( L_e \) 
is real and positive, and
the exponential \( \exp[{} - L_e(\phi,\dot \phi) ] \)
is a probability distribution
well-suited to Monte Carlo methods.
\par
This procedure is straightforward when 
the action is quadratic
in its time derivatives, 
and the integrals over the momenta are gaussian.
But when the equations
(\ref {def of pi}) that define
the momenta have square roots,
the hamiltonian usually has a square root.
When those equations are cubic
or quartic, the Legendre transformation
and the hamiltonian are complicated.
When they are worse than
quartic, no algebraic solution exists, and the 
hamiltonian typically is unknown. 
We show how to make path integrals
for such very awkward actions in 
Sec.~\ref{Path integrals for very awkward actions}\@.

\section {Path integrals for very awkward actions
\label {Path integrals for very awkward actions} }

\par
Our solution to the problem
of making a path integral without a
hamiltonian is to use
delta functionals to
impose Legendre's relation (\ref {def of pi})
between momenta and 
the fields and their derivatives.
Our formula for the partition function 
is a double path integral over the fields \( \phi \)
and over auxiliary time derivatives 
\( \dot \psi \)
\begin{equation}
   \begin{split}
Z(\beta) = {}&
\! \int \exp \bigg\{ \int_0^\beta \! \int 
\Big[ ( i \dot \phi_\ell
- \dot \psi_\ell ) \frac{ \p  L(\phi, \dot \psi) }
{ \p \dot \psi_\ell }
+ L(\phi, \dot \psi) 
\Big] dt d^3x \bigg\}  \lt| \det \! 
\lt( \frac{ \p^2  L(\phi, \dot \psi) }
{ \p \dot \psi_k \p \dot \psi_\ell }  \rt) \rt|
\, D\phi D\dot\psi 
\label {proposed path integral}
   \end{split}
\end{equation}
in which the \( n \times n \)
determinant converts 
\( D\dot \psi \) into \( D\pi \),
and the energy density
\begin{equation}
E(\phi, \dot \psi) = {}
\dot \psi_\ell \, \frac{ \p  L(\phi, \dot \psi) }
{ \p \dot \psi_\ell }
- L(\phi, \dot \psi) 
\label {pseudo-hamiltonian}
\end{equation}
is the hamiltonian density when
the time derivatives
\( \dot \psi_\ell  \) 
respect Legendre's relation (\ref {def of pi})\@.
If the action is time independent, then
the spatial integral of \( E(\phi, \dot \psi) \)
is a constant when 
\( \dot \psi_\ell = \dot \phi_\ell (\phi, \pi) \), and
the equations of motion are obeyed.
\par
The double path integral (\ref {proposed path integral})
for the partition function \( Z(\beta) \) is 
complex and ill-suited to estimation
by Monte Carlo methods.
We solve this problem in 
section~\ref {The Atlantic City method}\@.
\par
To derive our formula (\ref {proposed path integral}),
we write the path integral 
(\ref {leads to the partition function}) as
\begin{equation}
   \begin{split}
Z(\beta) = {}&
\int \exp \lt\{ \int_0^\beta \int \lt[ i \, \dot \phi_j \pi_j 
- ( \pi_k \dot \psi_k - L(\phi, \dot \psi) )
\rt] dt d^3x \rt\} \\
{}& \times \exp \lt[ i \int \lt( \pi_\ell 
- \frac{ \p  L }{ \p \dot \psi_\ell } \rt)
a_\ell \, d^4x \rt] 
\lt| \det \lt( \frac{ \p^2  L }
 { \p \dot \psi_k \p \dot \psi_\ell }  \rt) \rt| 
D\phi D\pi D\dot\psi Da 
\label {quartic path integral}
   \end{split}
\end{equation}
in which the integration over
the \( n \) auxiliary fields \( a_\ell \) makes
the second exponential 
a delta functional 
\( \d[ \pi -  \p L/ \p \dot \psi ] \) that enforces
the definition (\ref {def of pi}) of the momentum
\( \pi_j \)
as the derivative of the action density \( L \)
with respect to the time derivative \( \dot \phi_j \)\@.
The jacobian is an \( n \times n \) 
determinant that converts \( D\dot \psi \)
to \( D\pi \)\@.
The integration is over all fields 
that are periodic with period \( \beta \)\@.
Integrating first over \( a \), we get
\begin{equation}
   \begin{split}
Z(\beta) = {}&
\int \exp \lt\{ \int_0^\beta \int  \lt[ i \dot \phi_j \pi_j 
- ( \pi_k \dot \psi_k - L(\phi, \dot \psi) )
\rt] dt d^3x \rt\} \\
{}& \times \left\{ \prod_{\ell=1}^n \delta \lt[ \pi_\ell 
- \frac{ \p  L }{ \p \dot \psi_\ell } \rt] \rt\}
\lt| \det \lt( \frac{ \p^2  L }
{ \p \dot \psi_k \p \dot \psi_\ell }  \rt) \rt|
D\phi D\pi D\dot\psi  .
\label {triple path integral}
   \end{split}
\end{equation}
To integrate over the 
auxiliary time derivatives \( \dot \psi \),
we recall the delta-function rule
that if a vector 
\( g(x) = (g_1(x_1, \dots, x_n),   \dots, g_n(x_1, \dots, x_n)) \)
is zero only at \( x = x^0 \), then 
\begin{equation}
\d^n( g_1(x_1, \dots, x_n),   \dots, g_n(x_1, \dots, x_n)) 
\lt| \det\lt(\frac{\p g_k(x)}{\p x_\ell} \rt) \rt|  
= {}
\delta^n(x_1-x^0_1,\dots, x_n - x^0_n) .
\end{equation}
Thus integrating the triple path integral
(\ref {triple path integral})
over \( \dot \psi \), we find that
the delta functional
and the jacobian require 
the time derivatives to assume the values
\( \dot \psi_0(\phi,\pi) = \dot \phi(\phi,\pi) \)
that satisfy the definition (\ref {def of pi})
of the momenta,
and we get 
the path integral (\ref {leads to the partition function})
over \( \phi \) and \( \pi \)
\begin{equation}
   \begin{split}
Z(\beta) = {}&
\int \exp \lt\{ \int_0^\beta \int  \lt[ i \dot \phi \pi 
- ( \pi \dot \psi_0(\phi,\pi) - L(\phi, \dot \psi_0(\phi,\pi)) )
\rt] dt d^3x \rt\} D\phi D\pi 
\\
= {}&
\int \exp \lt\{ \int_0^\beta \int \left[ i \dot \phi_j \pi_j 
- H(\phi, \pi) \right] dt \, d^3x \rt\} D\phi D\pi .
\label {phi pi path integral redux}
   \end{split}
\end{equation}
On the other hand, if we integrate
the triple path integral (\ref {triple path integral})
over \( \pi \), then we get our proposed
formula (\ref {proposed path integral})
\begin{equation}
   \begin{split}
Z(\beta) = {}&
\! \int \exp \lt\{ \int_0^\beta \! \int \left[
( i \dot \phi_\ell - \dot \psi_\ell ) 
\frac{ \p  L(\phi, \dot \psi) }
{ \p \dot \psi_\ell }
+ L(\phi, \dot \psi) \right]
dt d^3x \rt\}
\lt| \det \lt( \frac{ \p^2  L(\phi, \dot \psi) }
{ \p \dot \psi_k \p \dot \psi_\ell }  \rt) \rt|
D\phi D\dot\psi .
\label {the path integral}
   \end{split}
\end{equation}
This functional integral generalizes 
the path integral to theories of scalar fields
in which the hamiltonian is unknown.
A similar formula should work in
theories of vector and tensor fields,
apart from the issue of constraints.
\par
Our first example is a free scalar field
with action density
\begin{equation}
L = - \half \, \p_\mu \phi \, \p^\mu \phi 
- \half m^2 \phi^2 .
\label {free L}
\end{equation}
The determinant in our formula
(\ref {the path integral}) is unity because
\begin{equation}
\frac{ \p^2  L(\phi, \dot \psi) }
{ \p \dot \psi^2 } = 1 .
\end{equation}
Using the abbreviation
\begin{equation}
\int_0^\beta dt \! \int d^3x \, \equiv \int^\beta d^4x,
\label {we will use the compact notation}
\end{equation}
we see that the proposed path integral
(\ref {the path integral}) 
for the free field theory (\ref {free L}) is
\begin{equation}
   \begin{split}
Z(\beta) = {}&
\int \exp \lt\{ \int^\beta \lt[  L(\phi, \dot \psi) 
+ \dot \psi ( i \dot \phi - \dot \psi )
\rt] d^4x \rt\}  D\phi D\dot \psi \\
= {}&
\int \exp \lt\{ \int^\beta \lt[  
\half \dot \psi^2 - \half (\grad \phi)^2 
- \half m^2 \phi^2 
+ \dot \psi ( i \dot \phi - \dot \psi )
\rt] d^4x \rt\}  D\phi D\dot \psi \\
= {}&
\int \exp \lt\{ \int^\beta \lt[ {} 
- \half \left( \dot \psi - i \dot \phi \right)^2
- \half \dot \phi^2 
- \half (\grad \phi)^2 
- \half m^2 \phi^2 
\rt] d^4x \rt\}  D\phi D\dot \psi \\
= {}&
\int \exp \lt\{ \int^\beta \lt[ {} - \half \dot \phi^2
- \half (\grad \phi)^2 
- \half m^2 \phi^2 
\rt] d^4x \rt\}  D\phi \\
= {}&\int \exp \lt[ {} - \int^\beta \!\!
L_e(\phi, \dot \phi) \, d^4x \rt]  D\phi 
\label {free path integral}
   \end{split}
\end{equation}
the standard result.

\par
Our second example is the scalar
Born-Infeld 
theory~\cite{Born:1934gh, *Born:1934dia, *Born:1935ap} 
with action density
\begin{equation}
L = {} M^4 \left( 1 - 
\sqrt{ 1-  M^{-4}
\lt(\dot \phi^2 
- (\grad \phi)^2 - m^2 \phi^2 \rt) }  \right)
\label {Lsqrt}
\end{equation}
and conjugate momentum 
\begin{equation}
\pi = \frac{\p L(\phi, \dot \phi)}{\p \dot \phi}
= \frac{\dot \phi}
{\sqrt{ 1-  M^{-4} \lt(\dot \phi^2 
- (\grad \phi)^2 - m^2 \phi^2 \rt) }} .
\end{equation}
The proposed path integral 
(\ref {the path integral}) is
\begin{equation}
   \begin{split}
Z(\beta)
= {}&  \int \exp \lt\{ \int^\beta \lt[  ( i \dot \phi
- \dot \psi ) \frac{ \p  L(\phi, \dot \psi)  }{ \p \dot \psi }
+ L(\phi, \dot \psi) 
\rt] d^4x \rt\}  
\lt| \frac{ \p^2  L(\phi, \dot \psi) }
{ \p \dot \psi^2  }  \rt|
 D\phi D\dot\psi
 \label {first Born-Infeld path integral}
   \end{split}
\end{equation}
in which 
\begin{equation}
\frac{ \p  L(\phi, \dot \psi)  }{ \p \dot \psi } = 
\frac{\dot \psi}
{\sqrt{ 1-  M^{-4} \lt(\dot \psi^2 
- (\grad \phi)^2 - m^2 \phi^2 \rt) }}
\label {first equation to invert}
\end{equation}
and 
\begin{equation}
\frac{ \p^2  L(\phi, \dot \psi) }{ \p \dot \psi^2  }
={} \frac{1 + M^{-4}\lt((\grad \phi)^2 + m^2 \phi^2\rt)}
{\lt[ 1-  M^{-4} \lt(\dot \psi^2 
- (\grad \phi)^2 - m^2 \phi^2 \rt) \rt]^{3/2} } .
\label {second equation to invert}
\end{equation}
Substituting these formulas into
(\ref {first Born-Infeld path integral}) gives
\begin{equation}
   \begin{split}
Z(\beta)  
= {}& \int \exp \Bigg\{ \int^\beta \Bigg[
\frac{( i \dot \phi - \dot \psi ) \dot \psi}
{\sqrt{ 1-  M^{-4} \lt(\dot \psi^2 
- (\grad \phi)^2 - m^2 \phi^2 \rt) } } 
\\
{}&  {} + M^4 \left( 1 - 
\sqrt{ 1-  M^{-4}
\lt(\dot \psi^2 
- (\grad \phi)^2 - m^2 \phi^2 \rt) } \right)
\Bigg] d^4x \Bigg\}  
\\
{}& \times
\frac{1 + M^{-4}\lt((\grad \phi)^2 + m^2 \phi^2\rt)}
{\lt[ 1-  M^{-4} \lt(\dot \psi^2 
- (\grad \phi)^2 - m^2 \phi^2 \rt) \rt]^{3/2} }
\, D\phi D\dot\psi .
\label {second Born-Infeld path integral}
   \end{split}
\end{equation}
\par
We can set
\begin{equation}
\pi ={} \frac{ \p  L(\phi, \dot \psi)  }{ \p \dot \psi } = 
\frac{\dot \psi}
{\sqrt{ 1-  M^{-4} \lt(\dot \psi^2 
- (\grad \phi)^2 - m^2 \phi^2 \rt) }}
\label {Setting pi =}
\end{equation}
and so absorb the jacobian in
\begin{equation}
d\pi ={} \frac{\p \pi}{\p \dot \psi} \, d \dot \psi
= \frac{ \p^2  L(\phi, \dot \psi) }{ \p \dot \psi^2  } \, d \dot \psi
={} \frac{1 + M^{-4}\lt((\grad \phi)^2 + m^2 \phi^2\rt)}
{\lt[ 1-  M^{-4} \lt(\dot \psi^2 
- (\grad \phi)^2 - m^2 \phi^2 \rt) \rt]^{3/2} } 
\, d \dot \psi  .
\label {we have}
\end{equation}
The partition function 
(\ref {second Born-Infeld path integral}) 
then is
\begin{equation}
   \begin{split}
Z(\beta)  
= {}& \int \exp \Bigg[ \int^\beta 
( i \dot \phi - \dot \psi ) \pi
 + M^4 \left( 1 - 
\sqrt{ 1-  
\lt(\dot \psi^2 
- (\grad \phi)^2 - m^2 \phi^2 \rt)/M^4 } \right)
d^4x \Bigg]  D\phi D \pi 
\label {third Born-Infeld path integral}
   \end{split}
\end{equation}
where now \( \dot \psi(\phi,\pi) \) 
is the function of \( \phi \) and \( \pi \)
defined by (\ref {Setting pi =})\@.
\par
The action of this theory is awkward, but not very awkward.
We can solve Legendre's equation
(\ref {Setting pi =})
for the time derivative \( \dot \phi \)
\begin{equation}
\dot \phi = \frac{\pi}{\sqrt{1 + M^{-4} \, \pi^2}}
\sqrt{1 + M^{-4} 
\lt( (\grad \phi)^2 + m^2 \phi^2 \rt)}  
\label {dot psi}
\end{equation}
and find as the hamiltonian density
\begin{equation}
   \begin{split}
H(\phi,\pi) = {}& \pi \dot \phi - L(\phi, \dot \phi) 
\\
={}& \frac{\pi^2\sqrt{1 + M^{-4} 
\lt( (\grad \phi)^2 + m^2 \phi^2 \rt)}}
{\sqrt{1 + M^{-4} \, \pi^2}} - M^4
\\
{}& + M^4
\sqrt{ 1-  M^{-4}
\lt( \frac{\pi^2 \left(1 + M^{-4} 
\lt( (\grad \phi)^2 + m^2 \phi^2 \rt)\right)}{1 + M^{-4} \, \pi^2}
- (\grad \phi)^2 - m^2 \phi^2 \rt) } 
\\
={}&  \frac{\pi^2\sqrt{M^4 +  (\grad \phi)^2 + m^2 \phi^2 }}
{\sqrt{M^4 + \pi^2}}
+ M^4 \frac{\sqrt{M^4 + (\grad \phi)^2 + m^2 \phi^2}}
{\sqrt{M^4 + \pi^2}} - M^4
\\
= {}& \sqrt{\lt(M^4 + \pi^2\rt)
\lt(M^4 + (\grad \phi)^2 + m^2 \phi^2\rt)}  - M^4 .
\label {H1}
   \end{split}
\end{equation}
Thus for this theory,
the double path integral
(\ref {the path integral for the partition function}) is
\begin{equation}
   \begin{split}
Z(\beta)
={}& \int \exp \lt\{ \int^\beta \lt[ i \dot \phi \pi  {} 
- \sqrt{\lt(M^4 + \pi^2\rt)\lt(M^4 
+ (\grad \phi)^2 + m^2 \phi^2\rt)} + M^4
\rt] d^4x \rt\} D\phi D\pi .
\label {Born-Infeld path integral}
   \end{split}
\end{equation}

\par
Our third example is the theory
defined by the action density
\begin{equation}
L = M^4 \exp( L_0/M^4 )
\label {exponential action}
\end{equation}
in which \( L_0 \) is the action
density (\ref {free L}) of the free field.
The derivatives
of \( L \) are
\begin{equation}
\frac{\p L}{\p \dot \psi} = M^{-4} \dot \psi \, L
\quad \mbox{and} \quad
\frac{\p^2 L}{\p \dot \psi^2}
= M^{-4} ( 1 + M^{-4} \dot \psi^2 ) \, L .
\end{equation}
So the proposed path integral is
\begin{equation}
\begin{split}
Z(\beta) = {}&
\int \exp \lt\{ \int^\beta \lt[ L(\phi, \dot \psi) 
+ \frac{ \p  L(\phi, \dot \psi) }{ \p \dot \psi } 
( i \dot \phi - \dot \psi )
\rt] d^4x \rt\} 
\lt| \frac{\p^2 L(\phi, \dot \psi) }
{\p \dot \psi^2} \rt| 
D\phi D\dot\psi  \\
={}& \int \exp \lt\{ \int^\beta   \lt[ 1
+ \frac{ \dot \psi 
( i \dot \phi - \dot \psi ) }
{M^4}
\rt] L(\phi, \dot \psi) \, d^4x \rt\} 
M^{-4} ( 1 + M^{-4} \dot \psi^2 ) \, L
 \,
D\phi D\dot\psi .
\label {basic path integral pi psi exponential}
\end{split}
\end{equation}
\par
Our fourth example is 
the Nambu-Got{\={o}} 
action density
\beq
L = {} - T_0 \,
\sqrt{ \lt( \dot X \cdot X' \rt)^2
- \lt( \dot X \rt)^2 \lt( X' \rt)^2 }
\label {Nambu Goto L}
\eeq
in which the tau or time derivatives
of the coordinate fields \( X^\mu \)
do not occur quadratically~\cite{CahillXIXi}.
The momenta are
\beq
\mathcal{P}^\tau_\mu = 
\frac{\partial L}
{\partial \dot X^\mu} = {}
- T_0 \,
\frac{(\dot X \cdot X') X'_\mu 
- (X')^2 \dot X_\mu}
{\sqrt{ \lt( \dot X \cdot X' \rt)^2
- \lt( \dot X \rt)^2 \lt( X' \rt)^2 }}
\label {Ptaumu}
\eeq
and the second derivatives 
of the Lagrange density are
\begin{equation}
   \begin{split}
\frac{\p^2 L}{\p \dot X^\mu \p \dot X^\nu}
= {}& T_0 \, \lt[
\frac{\eta_{\mu \nu} X'^2 - X'_\mu X'_\nu}
{ \sqrt{ \lt( \dot X \cdot X' \rt)^2
- \lt( \dot X \rt)^2 \lt( X' \rt)^2 } } \rt. \\
{}& \lt.
- \frac{ \lt( (\dot X \cdot X') X'_\mu 
- (X')^2 \dot X_\mu \rt)
\lt( (\dot X \cdot X') X'_\nu 
- (X')^2 \dot X_\nu \rt) }
{ \lt[ \lt( \dot X \cdot X' \rt)^2
- \lt( \dot X \rt)^2 \lt( X' \rt)^2 \rt]^{3/2} }
\rt]  .
\label {NG 2d derivatives}
   \end{split}
\end{equation}
The proposed partition function
(\ref {the path integral}) for
the Nambu-Got{\={o}} action is then
\begin{equation}
   \begin{split}
Z(\beta) = {}&
\! \! \int \! \exp \! \lt\{\int^\beta \lt[
( i \dot X^\mu
- \dot Y^\mu ) \frac{ \p  L(X, \dot Y) }
{ \p \dot Y^\mu }
+ L(X, \dot Y) \rt]
d\sigma d\tau \rt\}   
\lt| \det \! \lt[ \frac{ \p^2  L(X, \dot Y) }
{ \p \dot Y^\mu \p \dot Y^\nu }  \rt] \rt|
DX D\dot Y  
\label {the path integral with chi's for NG}
   \end{split}
\end{equation}
in which the formulas (\ref {Ptaumu})
and (\ref {NG 2d derivatives}) 
(with \( \dot X^\mu \to \dot Y^\mu \)) are to be
substituted for the first and second
derivatives of the action density \( L \)
with respect to the tau derivatives
\( \dot Y^\mu \)\@.
But because the action density
\( L \) is a homogeneous function
of degree 1
of the time (and space) derivatives
of the fields \( X^\mu \),
its energy density vanishes
independently of the equations of motion 
\begin{equation}
E ={} \dot X^\mu 
\frac{\p L}{\dot X^\mu} - L = 0
\label {NG energy vanishes}
\end{equation}
by Euler's theorem.
Thus the partition function
(\ref {the path integral with chi's for NG})
is simply
\begin{equation}
   \begin{split}
Z(\beta) = {}&
\! \! \int \! \exp \! \lt[\int^\beta 
i \dot X^\mu
 \frac{ \p  L(X, \dot Y) }
{ \p \dot Y^\mu }
d\sigma d\tau \rt]   
\lt| \det \! \lt[ \frac{ \p^2  L(X, \dot Y) }
{ \p \dot Y^\mu \p \dot Y^\nu }  \rt] \rt|
DX D\dot Y  .
\label {the NG path integral simply}
   \end{split}
\end{equation}
But since we know that the hamiltonian
(\ref {NG energy vanishes}) vanishes,
we can use the simpler formula
(\ref {leads to the partition function})
and get 
for the partition function
the badly divergent expression
\begin{equation}
   \begin{split}
Z(\beta) = {}&
\! \! \int \! \exp \! \lt[\int^\beta 
i \dot X^\mu \mathcal{P}^\tau_\mu 
\, d\sigma d\tau \rt]
DX D\mathcal{P} .
\label {simpler formula for string}
   \end{split}
\end{equation}

\section{The Atlantic City method
\label {The Atlantic City method}}

Monte Carlos let us estimate the mean values
of observables weighted by probability 
distributions~\cite{CahillXIVi}.
They fail when the weight function assumes
negative or complex values.
This failure is one aspect of the sign 
problem.
The double-ratio 
trick (\ref{observable}--\ref{first probability distribution})
outlined in the
appendix~\ref {Ratios of complex Monte Carlos are unreliable}
is unreliable.
\par
These problems are not hopeless however.
For although the weight functions 
of the double path integrals
(\ref{the path integral for the partition function}) and 
(\ref {first proposed path integral})
are complex,
the integrals of these 
complex weight functions over 
the momenta \( \pi \) or over 
the auxiliary time derivatives \( \dot \psi \)
are real and positive.
They are the probability distribution
that determines the partition function
and the mean values of observables.
\par
If one can't do these
integrals analytically,
one can do them numerically.
These numerical integrations
are well suited to parallel computation.
In the Atlantic City method,
one numerically integrates in parallel over 
the momenta \( \pi \) or over 
the auxiliary time derivatives \( \dot \psi \)
in the double path integrals
(\ref{the path integral for the partition function} or 
\ref {first proposed path integral}) 
and stores the values
of these integrals in a lookup table.
One then uses the Monte Carlo
method guided by the
stored integrals 
to estimate the mean values of observables.
\par
Our main goal is
to study the ground states
of field theories, but 
for simplicity in this paper
we will explain and test 
the Atlantic City method in the context 
of quantum mechanics.
\par
If the action is awkward,
but not very awkward,  
then we can find the hamiltonian
\( H(q,p) \) but can't integrate analytically
over the momentum \( p \)\@.
Then the partition function \( Z(\beta) \) is
\begin{equation} 
   \begin{split}
 Z(\beta) ={}& \tr \, e^{-\beta H} 
=  \int \exp\left\{ \int_0^\beta \left[ i \dot q p - H(q,p) \right] dt \right\}
Dp Dq .
 \label {The partition function then is}
    \end{split}
\end{equation}
We use the approximation
\begin{equation}
\begin{split}
\la q_{\ell+1} | e^{-a H(q, p)} | q_\ell \ra
\approx {}& 
\int dp \, \la q_{\ell+1} | p\ra \la p |  e^{-a H(q_\ell, p)} | q_\ell \ra
\\
= {}& 
\int_{-\infty}^\infty \frac{dp}{\sqrt{2\pi}} \,
\exp \left[i (q_{\ell+1} - q_\ell) p \, 
- a \,  H(q_\ell, p) \right] 
\label {using the approximation}
\end{split}
\end{equation}
to estimate the partition function as
the multiple integral
\begin{equation}
\begin{split}
Z(\beta)={}& \prod_{j=1}^{n} \int \frac{dp_j dq_j}{2\pi} 
 \exp \left[  i (q_{j+1} - q_j) p_j 
 - a H(q_j,p_j) \right]   
 \label {The partition function then is}
    \end{split}
\end{equation}
in which \( n = \beta/a \), and 
the paths are periodic \( q_{n+1} = q_1 \)\@.
\par
The path integral
\begin{equation}   
P[q,\beta] = {} \int \exp
\left\{ \int_0^\beta \left[ i \dot q p - H(q,p) \right] dt \right\}
Dp
\label {We know that }
\end{equation}
is an unnormalized functional probability distribution
that assigns a number \( P[q, \beta] \)
to every path \( q(t) \)\@.
It is the limit as \( n \to \infty \) and \(  a =\beta/n \to 0 \)
of the multiple integral
\begin{equation}   
P_n[q,\beta] ={} \prod_{j=1}^n \int\frac{dp_j }{\sqrt{2\pi}} 
 \exp \left[  i (q_{j+1} - q_j) p_j 
  - a H(q_j,p_j) \right] .
 \label {In the limit }
\end{equation}
If the hamiltonian is even in the momentum,
then this probability distribution is real
\begin{equation}   
P_n[q,\beta] ={} \prod_{j=1}^n \int\frac{dp_j }{\sqrt{2\pi}} 
 \cos[(q_{j+1} - q_j) p_j] 
 e^{ -a H(q_j,p_j) } .
 \label {P when H is p even}
\end{equation}
The partition function \( Z(\beta) \) is
\begin{equation}
   \begin{split}
      Z(\beta) = {}&  \int P[q, \beta] \, Dq
     = {} \int \prod_{j=1}^n \frac{dq_j}{\sqrt{2\pi}} \,
      P_n[q,\beta] .
  \label {I hope The partition function is}   
   \end{split}
\end{equation}
The mean value of the energy 
at inverse temperature \( \beta \) is
\begin{equation}
   \begin{split}
\la H \ra_\beta ={}
\frac{\tr \, H \, e^{- \beta H}} {\tr \, e^{- \beta H}}  
={}& - \frac{1}{Z(\beta)} \frac{d Z(\beta)}{d \beta}
= - \frac{1}{Z(\beta)} \int \prod_{j=1}^n \frac{dq_j}{\sqrt{2\pi}} \,
\frac{d P_n[q,\beta] }{d \beta} .
\label {mean value of the energy at beta}
   \end{split}
\end{equation}
The derivative of the probability distribution 
with respect to \( \beta = n \, a \) is
\begin{equation}
 \begin{split}
{} - \frac{d  P_n[q,\beta] }{d \beta}
={}& \frac{1}{n} \sum_{k=1}^n 
\int \prod_{j=1}^n \frac{dp_j }{\sqrt{2\pi}} \, H(q_k,p_k) 
\, e^{ i (q_{j+1} - q_j) p_j - a H(q_j,p_j) } .
   \end{split}
\end{equation}
So the mean value of the hamiltonian 
at inverse temperature \( \beta \) is
\begin{equation}
   \begin{split}
\la H \ra_\beta ={}& 
\frac{1}{n} \sum_{k=1}^n 
\int \prod_{j=1}^n dq_j \, dp_j \,
\, H(q_k,p_k) 
\, e^{ i (q_{j+1} - q_j) p_j  - a H(q_j,p_j) }
\\
{}& \left/ \int \prod_{j=1}^n  
dq_j \, dp_j \,
e^{ i (q_{j+1} - q_j) p_j  - a H(q_j,p_j) } \right. .
 \label {maybe big eq for E0}
   \end{split}
\end{equation}
\par
In the Atlantic City method,
one does the \( p \) integrations numerically, setting
\begin{equation}
\begin{split}
A(q_{\ell+1}, q_\ell) 
= \int_{-\infty}^\infty \frac{dp}{\sqrt{2\pi}} \,
\exp \left[i (q_{\ell+1} - q_\ell) p \, 
- a \,  H(q_\ell, p)\right] 
\label {integration numerically, we define}
\end{split}
\end{equation}
and
\begin{equation}
C(q_{\ell+1},q_\ell) = {}  \int_{-\infty}^\infty 
\frac{dp}{\sqrt{2\pi}} \,
H(q_\ell, p) \,
\exp \left[i (q_{\ell+1} - q_\ell) p \, 
- a \,  H(q_\ell, p) \right] .
\label {C(q',q)}
\end{equation}
If one uses \( N \) values of \( q_\ell \),
then one does these \( 2 N^2 \)
numerical integrals.
One may do them in parallel.
\par
In most problems of interest,
the hamiltonian is an even function
of the momentum, \( H(q,-p) = H(q,p) \),
and the integrals 
(\ref {integration numerically, we define} \&
\ref {C(q',q)}) are real
\begin{equation}
\begin{split}
A(q_{\ell+1}, q_\ell) = {} & \sqrt{\frac{2}{\pi}}
\int_0^\infty dp \,
\cos \left[(q_{\ell+1} - q_\ell) \, p \right] \,
e^{- a \, H(q_\ell, p) } 
\\
C(q_{\ell+1}, q_\ell) = {} & \sqrt{\frac{2}{\pi}}
\int_0^\infty dp \, H(q_\ell, p) \,
\cos \left[ (q_{\ell+1} - q_\ell) \, p \right] \,
e^{- a \,  H(q_\ell, p) }  .
\label {and the integrals are real}
\end{split}
\end{equation}
\par
One's tables need run only over
\( q_{\ell+1} \ge q_\ell \)\@.
We have found it convenient
to use the variables 
\( dq_\ell = | q_{\ell+1} - q_\ell | \)
and \( q_\ell \) and to adjust 
the resolution of the tables according
to the variation of the integrals
\( A(dq_\ell, q_\ell) \) and 
\( C(dq_\ell, q_\ell) \)\@.
\par
In terms of these numerical integrals,
the mean value of the hamiltonian is
\begin{equation}
   \begin{split}
\la H \ra_\beta ={}&
   \frac{1}{n} \sum_{k=1}^n 
\int dq_k \prod_{j=1, \, j \ne k}^n \! dq_j  \,
C(q_{k+1}, q_k) \, A(q_{j+1},q_j)  
 \left/ \int \prod_{j=1}^n  
dq_j \, A(q_{j+1},q_j) \right. 
 \label {medium eq for E0}
   \end{split}
\end{equation}
which we may write as
\begin{equation}
   \begin{split}
\la H \ra_\beta ={}&
   \frac{1}{n} \sum_{k=1}^n \int 
   \prod_{j=1}^n \! dq_j  \,
\frac{C(q_{k+1}, q_k)}{  A(q_{k+1},q_k) }
\, A(q_{j+1},q_j) 
 \left/ \int \prod_{j=1}^n  
dq_j \, A(q_{j+1},q_j) \right. .
 \label {medium eq for E0}
   \end{split}
\end{equation}
We do a Monte Carlo over the 
probability distribution
\begin{equation}
P(q) ={} \left.
\prod_{j=1}^n  A(q_{j+1},q_j) 
\right/ \int \prod_{j=1}^n  
dq_j \, A(q_{j+1},q_j) 
\label {a Monte Carlo over the probability distribution}
\end{equation}
and measure the ratio
\begin{equation}
\la H \ra_\beta ={}
\left\la \frac{1}{n} \sum_{k=1}^n 
\frac{C(q_{k+1}, q_k)}{  A(q_{k+1},q_k) }
\right\ra = \int \frac{1}{n} \sum_{k=1}^n 
\frac{C(q_{k+1}, q_k)}{  A(q_{k+1},q_k) }
\, P(q) \, Dq .
\label {and measure} 
\end{equation}
\par
When the hamiltonian is a monotonically
increasing, even function of the momentum,
the integration (\ref{and the integrals are real})
for \( A(q_{\ell + 1}, q_\ell) \) is positive
over every interval
\begin{equation}
\frac{2\pi n}{\Delta q_\ell} 
\, \le \, {} p \, \le {} 
\frac{2\pi (n+1)}{\Delta q_\ell}
\end{equation}
for \( n = 0, 1, 2 \dots \)
where \( \Delta q_\ell = q_{\ell+1} - q_\ell \)\@.
The reason is that 
when \( H(q_\ell, p) \) increases with \( p \),
the positive
integral from \( p = 2\pi n/\Delta q_\ell \)
to \( p = 2\pi (n + \thalf)/ \Delta q_\ell \)
weighted by \( \exp( - a H(q_\ell, p) ) \)
with \( p \) in that interval
exceeds the negative 
integral from \( p = 2\pi (n + \thalf)/ \Delta q_\ell \)
to \( p = 2\pi (n + 1)/ \Delta q_\ell \)
weighted by \( \exp( - a H(q_\ell, p) ) \)
with \( p \) in this second interval.
\par
Thus as long as the hamiltonian is a monotonically
increasing, even function of the momentum,
the product 
\( A(q_{j+1}, q'_j) \, A(q'_{j}, q_{j-1}) \)
will be an unnormalized probability
distribution in the variable \( q'_j \)\@.
It is a simple matter to
have one's Monte Carlo code
report the minimum 
value of the integral \( A(q_{\ell + 1}, q_\ell) \) 
(\ref {and the integrals are real})
and to check that it is positive.
\par
To take a Metropolis step,  
we pick a new \( q'_j \)
and look up the value of the
(unnormalized) probability distribution
\begin{equation}
P(q'_j)={}
 A(q_{j+1}, q'_j) \, A(q'_{j}, q_{j-1}) .
 \label {probability distribution 1}
\end{equation}
Usually, the random points \( q_{j+1} \),
\( q'_j \), and \( q_{j-1} \) are not be among
the \( q_j \)'s in our tables, so
our computers use a bilinear
interpolation to approximate
\( A(q_{j+1}, q'_j) \) and \( A(q'_{j}, q_{j-1}) \)\@.
\par
If \( P(q'_j) \ge P(q_j) \), then we accept
the new  \( q'_j \)\@.
If \( P(q'_j) < P(q_j) \), then we 
accept the new  \( q'_j \) with 
conditional probability
\begin{equation}
P(q_j \to q'_j) = P(q_j')/P(q_j) 
 \label {probability distribution 2}
\end{equation}
and otherwise reject it.

\section{Application of the Atlantic City method to the  Born-Infeld oscillator
\label {Application of the Atlantic City method to the 
Born-Infeld oscillator}}

In this section we demonstrate
and test our Atlantic City model
on a theory with an awkward action,
the quantum-mechanical version of
the scalar Born-Infeld model
(\ref {Lsqrt}--\ref {Born-Infeld path integral})\@.
The lagrangian of this model is
\begin{equation}
L = {} Mc^2 - Mc^2 \left[ 1 - \frac{m}{Mc^2} 
\left( \dot q^2 - \omega^2 q^2 \right) \right]^{1/2} .
\label {Born-Infeld qm L}
\end{equation}
The momentum is
\begin{equation}
p ={} \frac{m \dot q}
{\sqrt{1 - m\left( \dot q^2 - \omega^2 q^2 \right)/(Mc^2) }} ,
\label {The momentum is}
\end{equation}
and the velocity is
\begin{equation}
\dot q ={} \frac{p}{m} \frac{\sqrt{1 + m \omega^2 q^2 /Mc^2}}
{\sqrt{1 + p^2/(mMc^2)}} .
\label {velocity}
\end{equation}
The hamiltonian of the Born-Infeld 
oscillator is
\begin{equation}
H ={} \sqrt{ \left( p^2/m + Mc^2 \right) \,\,
\left( Mc^2 + m \omega^2 q^2 \right) } - Mc^2 .
\label {Born-Infeld hamiltonian}
\end{equation}
In terms of the hamiltonian
\( H_0 = p^2/2m + m \omega^2 q^2/2 \)
of the harmonic oscillator,
the hamiltonian \( H \) of the Born-Infeld 
oscillator in the limit \( M/m \gg 1 \) is
\begin{equation}
   \begin{split}
H ={}& H_0 - \frac{1}{8Mc^2} 
 \left(\frac{p^2}{2m} - \frac{m\omega^2 q^2}{2} \right)^2
 \left( 1 - \frac{H_0}{Mc^2} \right) + \dots
\label {When M >> m, the first few terms of the hamiltonian}
   \end{split}
\end{equation}
and
\begin{equation}
H={} \sqrt{Mc^2m\omega^2q^2}
\Big[1+\frac{p^2}{2mMc^2} -  \frac{p^4}{8(mMc^2)^2} \Big]
\Big[1 + \frac{Mc^2}{2m\omega^2 q^2} - \frac{(Mc^2)^2}{32(m\omega^2q^2)^2} \Big] + \dots
\end{equation}
for \( M/m \ll 1 \)\@.

\par
With \( \hbar = c = 1 \) and \( \beta = n \, a \),
the partition function is 
\begin{equation} 
   \begin{split}
 Z(\beta) ={}& \tr \, e^{-\beta H} 
=  \int \exp\left\{ \int_0^\beta \left[ i \dot q p - H(q,p) \right] dt \right\}
Dp Dq 
\\
\approx{}& \prod_{j=1}^n \int\frac{dp_j dq_j}{2\pi} 
 \exp \left[   i (q_{j+1} - q_j) p_j 
- a H(q_j,p_j) \right]  
 \\
\approx{}& \prod_{j=1}^n \int\frac{dp_j dq_j}{2\pi} 
 \exp \left[   i (q_{j+1} - q_j) p_j 
- a \left( \sqrt{ \left(p^2_j/m + M \right)
\left(M + m \omega^2 q^2_j \right)} - M \right)
 \right]  .
 \label {B-I partition function}
   \end{split}
\end{equation}
\par
In the limit \( M = 0 \),
the hamiltonian (\ref{Born-Infeld hamiltonian}) is
\( H_{M=0} = {} | \omega \, p \, q | \),
which is so simple 
that we can integrate
over the momentum and write
the partition function as
an ordinary path integral
\begin{equation} 
   \begin{split}
 Z(\beta) \approx {}& 
\prod_{j=1}^n \int\frac{dp_j dq_j}{2\pi} 
 \exp \left[   i (q_{j+1} - q_j) p_j 
- a | \omega \, p_j \, q_j |
 \right]  
 \\
 ={}& 
 \prod_{j=1}^n \int \frac{ dq_j}{\pi} 
 \frac{ a \, | \omega \, q_j |}
 {a^2 \, \omega^2 q_j^2 + (q_{j+1} - q_j)^2} .
 \label {B-I partition function M=0}
   \end{split}
\end{equation} 
The naive formula for the partition function
is to replace \( t \) by \( - i \beta \)
in the path integral for the amplitude
\begin{equation}
\la q(t) | e^{-itH} | q(0) \ra ={}
\int e^{i \int L dt} \, Dq .
\label{path integral for the amplitude}
\end{equation}
If we applied this rule to the 
action density (\ref {Born-Infeld qm L})
in the limit \( M \to 0 \)
keeping \( m M = 1 \),
then we'd get for the partition function
\begin{equation}
Z(\beta)_{\mbox{\footnotesize{naive}}} 
\approx {}
 \prod_{j=1}^n \int \frac{ dq_j}{2\pi} 
\, e^{- \sqrt{(q_{j+1} - q_j)^2 
 + a^2\omega^2 q_j^2}}
\end{equation}
which is very different from the
correct formula (\ref{B-I partition function M=0})\@.
\par
In terms of the variables to 
\( q' = q \sqrt{m} \) and \( p' = p/\sqrt{m} \),
which satisfy the commutation relation
\( [ q', p' ] = i \), the Born-Infeld hamiltonian
(\ref {Born-Infeld hamiltonian}) is
\begin{equation}
H ={} \sqrt{ \left( p'^2 + M \right) \,\,
\left( M + \omega^2 q'^2 \right) } - M ,
\label {Born-Infeld hamiltonian reveals}
\end{equation}
which shows that the energy levels
are independent of the mass parameter \( m \)\@.
To simplify our notation and expose the actual
dependence of these energies, we change
variables again to 
\( p' = \sqrt{M} p'' \) and
\( q' =  \sqrt{M} q'' / \omega \)\@.
After we drop all the primes,
we have
\begin{equation}
H ={} M \left[ \sqrt{ \left( p^2 + 1 \right) \,\,
\left( q^2 + 1 \right) } - 1 \right] ,
\label {Born-Infeld hamiltonian revealed again}
\end{equation}
and
\begin{equation} 
   \begin{split}
 Z(\beta) \approx {}& \prod_{j=1}^n \int
 \frac{Mdp_j dq_j}{2\pi \omega} 
 \exp \left\{  \left[
 i \frac{M}{\omega} (q_{j+1} - q_j) p_j 
- a M \left(\sqrt{\left(p^2_j + 1 \right) 
\left( q^2_j + 1\right)} - 1\right) \right]
 \right\}  .
 \label {B-I partition function}
   \end{split}
\end{equation}
The mean value of the hamiltonian 
at inverse temperature \( \beta = n a \) is
\begin{equation} 
   \begin{split}
\la H \ra_\beta ={}& 
- \frac{\tr \, H \, e^{- \beta H}} {\tr \, e^{- \beta H}}  
= {} - \frac{1}{Z(\beta)} \frac{d Z(\beta)}{d \beta} 
= {} - \frac{1}{n Z(\beta)} \frac{d Z(\beta)}{d a} .
\label {mean value of the hamiltonian }
   \end{split}
\end{equation}
\par
The energy \( \la H \ra_\beta \)
is a function of the ratio \( M/\omega \)
and is proportional to \( M \) 
\begin{equation} 
   \begin{split}
\la H \ra_\beta= {}& 
\prod_{j=1}^n \int dp_j dq_j 
\left[ \frac{M}{n}\sum_{\ell=0}^n  
\left(\sqrt{ \left( p^2_j + 1 \right)
\left( q^2_j + 1 \right) } - 1\right)
\right]
\\
{}& \times 
\exp \left[ 
 i \frac{M}{\omega} (q_{j+1} - q_j) p_j 
- a M \left(\sqrt{ \left( p^2_j + 1 \right)
\left( q^2_j + 1 \right) } - 1\right)
 \right]
 \\
 {}& \left/
\prod_{j=1}^n \int \! dp_j dq_j
\exp \left[ i \frac{M}{\omega} (q_{j+1} - q_j) p_j 
- a M \left(\sqrt{ \left( p^2_j + 1 \right)
\left( q^2_j + 1 \right) } - 1\right)
 \right] \right. .
 \label {B-I partition function}
   \end{split}
\end{equation}
The ground-state energy is the limit 
of the ratio as \( \beta \to \infty \)
and \( a \to 0 \)\@.
\par
We wrote Fortran 90 codes to compute
in parallel the momentum integrals
\begin{equation}
   \begin{split}
A(q_{\ell+1}, q_\ell) ={}& 
\int_0^\infty \! dp 
\cos \left[ \frac{M}{\omega}(q_{\ell+1} - q_\ell) p \right]
\exp \left[   
- a M \left(\sqrt{ \left( p^2 + 1 \right)
\left( q^2_\ell + 1 \right) } - 1\right)
 \right]
\label {Our first numerical integral is}
   \end{split}
\end{equation}
and
\begin{equation}
   \begin{split}
      C(q_{\ell+1}, q_\ell) = {}&
     M \int_0^\infty \! dp 
\left(\sqrt{ \left( p^2 + 1 \right)
\left( q^2_\ell + 1 \right) } - 1\right)     
\\ 
{}& \times      
\cos \left[ \frac{M}{\omega}(q_{\ell+1} - q_\ell) p \right]
\exp \left[   
- a M \left(\sqrt{ \left( p^2 + 1 \right)
\left( q^2_\ell + 1 \right) } - 1\right)
 \right]
\label {Our second numerical integral is}
         \end{split}
\end{equation}
for suitably large sets of values of \( q_\ell \)
and \( q_{\ell + 1} \)
and stored them in lookup tables.
We then used the lookup tables
in standard Monte Carlos
with a Metropolis step 
(\ref {probability distribution 1}--\ref 
{probability distribution 2}) 
to estimate the mean value
of the hamiltonian at inverse temperature \( \beta \)
\begin{equation}
\la H \ra_\beta ={}
\left\la \frac{1}{n} \sum_{k=1}^n 
\frac{C(q_{k+1}, q_k)}{  A(q_{k+1},q_k) }
\right\ra = \int \frac{1}{n} \sum_{k=1}^n 
\frac{C(q_{k+1}, q_k)}{  A(q_{k+1},q_k) }
\, P_n[q,\beta] \, Dq 
\label {and measure} 
\end{equation}
in which the unnormalized probability
distribution is
\begin{equation}
P_n(q,\beta) ={} \prod_{\ell=1}^n A(q_{\ell + 1}, q_\ell) 
\label {un-normalized probability}
\end{equation}
and \( q_{n+1} \equiv q_1 \)\@.
\par
\begin{figure}[H]
\begin{center}
\includegraphics[trim={0 2.75in 0 2.75in}, clip, 
width=\textwidth]{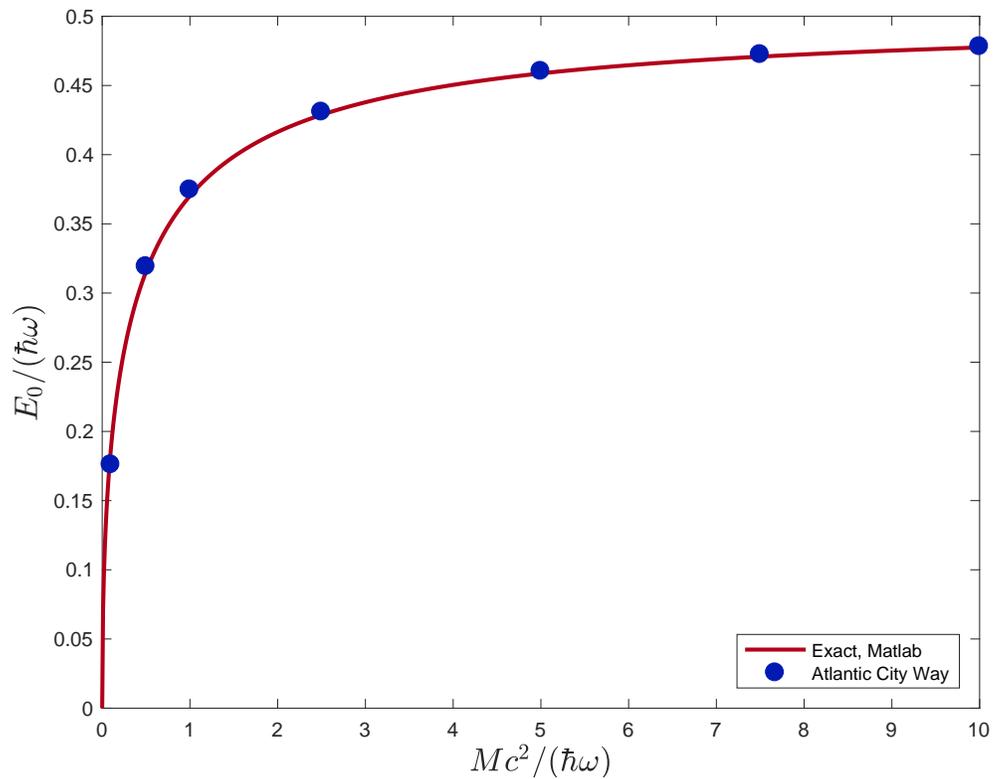}
\caption{Our Atlantic City  estimates 
(\ref {Our first numerical integral is}--\ref {and measure}, blue dots) 
of the ground-state energies \( E_0/(\hbar \omega) \)
of the Born-Infeld oscillator are plotted along 
with the exact values (Matlab, red curve) 
for \( 0.1 \le Mc^2/(\hbar \omega) \le 10 \),
\( a \omega = 0.1 \), and \( \beta = 10^3/M \)\@.}
\label {Born-Infeld big masses}
\end{center}
\end{figure}

The Monte Carlo codes run fast;
all the work is in the lookup tables. 
We made lookup
tables for \( 0.1 \le Mc^2/(\hbar \omega) \le 10 \),
\( a \omega = 0.1 \), and \( \beta = 10^3/M \)\@.
We plotted our Atlantic City   
(\ref {Our first numerical integral is}--\ref {and measure})
estimates of the ground-state
energy of the Born-Infeld oscillator
as blue dots
in Fig.\,\ref{Born-Infeld big masses}
and listed them in Table\,\ref{bigMasses}\@.
The integrals 
(\ref {Our first numerical integral is} \&
\ref {Our second numerical integral is})
have the exponential term
\( \exp [ {} - a M \sqrt{( p^2 + 1)
( q^2_\ell + 1 ) } ] \) and so
converge faster at big \( M \)
for fixed
\( a \) and \( \omega \)\@.
The statistical errors are smaller
than the dots.
\par
To test these results,
we used Matlab to compute the exact eigenvalues
of the Born-Infeld oscillator.
In terms of the harmonic-oscillator variables
\( a ={} \sqrt{m \omega/2} 
\left[ q + i p/(m \omega) \right] \)
and
\( a^\dagger ={} \sqrt{m \omega/2} 
\left[ q - i p/(m \omega) \right] \),
the operators \( q \) and \( p \) are
\( q = \left( a^\dagger + a \right)/\sqrt{2 m \omega} \)
and \( p = i \sqrt{m \omega/2} \left( a^\dagger - a \right) \),
and so
the hamiltonian
(\ref {Born-Infeld hamiltonian}) is
\begin{equation}
H ={} \sqrt{Mc^2 - \frac{\omega}{2}
\left( a^\dagger - a \right)^2}
\sqrt{Mc^2 + \frac{\omega}{2} 
\left( a^\dagger + a \right)^2} - Mc^2 
\label {Born-Infeld hamiltonian a a'}
\end{equation}
in which the mass \( m \) does
not appear.
We made a matrix
\( a \) as diag(sqrt([1:Nmax]),1) with Nmax = 1000
and \( a^\dagger \) as its transpose.
The Matlab command eig(sqrtm(H))
then gave the exact energy eigenvalues,
which generated the
red curves in the figures and the exact
results in the tables.

\par
\begin{table}[H]
\caption{Exact (Matlab) and 
Atlantic City  results 
(\ref {Our first numerical integral is}--\ref {and measure})
for the ground-state energy \( E_0/(\hbar \omega) \) 
of the Born-Infeld hamiltonian
(\ref{Born-Infeld hamiltonian})
for \( 0.1 \le Mc^2/(\hbar \omega) \le 10 \)\@.}
\begin{center}
\begin{tabular}{|c|c|c|}
\hline
\( Mc^2/(\hbar \omega) \)& \( E_0/(\hbar \omega) \) exact 
& \(E_0/(\hbar \omega)\)  Atlantic City \\
\hline
0.1&0.1881&0.1759\\
\hline
0.5&0.3155&0.3191\\
\hline
1.0&0.3702&0.3746\\
\hline
2.5&0.4288&0.4308\\
\hline
5.0&0.4587&0.4603\\
\hline
7.5&0.4708&0.4723\\
\hline
10.0&0.4774&0.4781\\
\hline
\end{tabular}
\end{center}
\label {bigMasses}
\end{table}

\section{The Atlantic City model applied to a very awkward action
\label{The Atlantic City model applied to a very awkward action}}

In this section,
we test our Atlantic City method
by using it to find the 
ground-state energy 
of the Born-Infeld oscillator
considered 
as a theory with a very awkward action.
That is, we pretend that
we don't know the Born-Infeld hamiltonian 
(\ref {Born-Infeld hamiltonian})
and use our Atlantic City method to
evaluate the complex path integral
(\ref {first proposed path integral})
for its partition function.
\par
Instead of the partition function
(\ref {B-I partition function}),
we have the partition function 
\begin{equation}
   \begin{split}
Z(\beta) ={}& \int \exp \lt\{ \int^\beta \lt[  ( i \dot q
- \dot s ) \frac{ \p  L(q, \dot s)  }{ \p \dot s }
+ L(q, \dot s) 
\rt] dt \rt\}  
\lt| \frac{ \p^2  L(q, \dot s) }
{ \p \dot s^2  }  \rt|
 Dq D\dot s
\\
={}& \int \exp\left\{
\int^\beta \left[
\frac{m i\dot q \dot s - m\omega^2 q^2 -M}
{\sqrt{1 - m \left( \dot s^2 - \omega^2 q^2 \right)/M}}
+ M \right] \, dt \right\}
\\
{}& \times \, \frac{m + m^2 \omega^2 q^2/M}
{\left[ 1 - 
m \left( \dot s ^2 - \omega^2 q^2 \right)/M\right]^{3/2}}
\, Dq D\dot s .
\label {instead of the partition function}
   \end{split}
\end{equation}
Sending \( q_j \to \sqrt{M/m} \, q_j/\omega \) and
\( \dot s_j \to \sqrt{M/m} \, \dot s_j \), 
we can write \( Z(\beta)  \) as
\begin{equation}
   \begin{split}
Z(\beta) ={}& \int \exp\left\{
\int^\beta \left[ 
\frac{i (M/\omega) \dot q \dot s - M q^2 - M}
{\sqrt{1 + q^2 - \dot s^2}}
+ M \right] \, dt \right\}
\frac{1 + q^2}
{\left[ 1 + q^2 - \dot s ^2 \right]^{3/2}}
\frac{M}{\omega} Dq D\dot s .
\label {we can write Z(beta) as}
   \end{split}
\end{equation}
\par
Setting \( dt = a \) and sending \( a \to a/M \), 
we approximate this path integral 
on an \( n \by n \) lattice
of spacing \( a = \beta/n \)
as the multiple integral
\begin{equation}
   \begin{split}
Z(\beta) ={}& \prod_{j=1}^n \int 
\frac{M d\dot s_j dq_j}{2\pi \omega} 
\exp  \left[ 
i \frac{M}{\omega} \left(
\frac{\left( q_{j+1}- q_j \right) \dot s_j}
{\sqrt{1 + q^2_j - \dot s^2_j }}
 \right) 
- a M \left(
\frac{q^2_j + 1}{\sqrt{1 + q^2_j - \dot s^2_j }} - 1 \right) 
\right] 
\\
{}&\quad \times \, 
\frac{ 1 +  q^2_j }
{\left[ 1 + q^2_j - \dot s ^2_j \right]^{3/2}} 
\label {We approximate this path integral}
   \end{split}
\end{equation}
in which the lower limits are \( q_j = \dot s_j = 0 \)
and the upper limits are \( q_j \to \infty \)
and \( \dot s_j \le \sqrt{ q^2_j + 1 } \)\@.
\par
Apart from the phase factor, the integrand
is even in \( \dot s \)\@.
We numerically compute the integrals
\begin{equation}
   \begin{split}
   A(q_{j+1}, q_j) ={}& \int_0^{\sqrt{ q^2_j + 1 }}  
   d\dot s \,
\cos \left[ \frac{M}{\omega} \left(
\frac{\left( q_{j+1}- q_j \right) \dot s_j}
{\sqrt{1 + q^2_j - \dot s^2_j }}
 \right) \right]
 \exp \left[ - a M\left(
\frac{q^2_j + 1}{\sqrt{1 + q^2_j - \dot s^2_j }} 
- 1 \right) \right] 
\\
{}&\quad \times \,\frac{ 1 +  q^2_j }
{\left[ 1 + q^2_j - \dot s ^2_j \right]^{3/2}}
\label {I L}
   \end{split}
\end{equation}
and
\begin{equation}
   \begin{split}
   C(q_{j+1}, q_j) ={}& M \int_0^{\sqrt{ q^2_j + 1 }} 
   d\dot s \,
   \left[
  \frac{q^2_j + 1}{\sqrt{1 + q^2_j - \dot s^2_j }}
- 1    \right]
\cos \left[ \frac{M}{\omega} \left(
\frac{\left( q_{j+1}- q_j \right) \dot s_j}
{\sqrt{1 + q^2_j - \dot s^2_j }}
 \right) \right]
 \\
 {}& \times 
 \exp \left[ - a M \left(
\frac{q^2_j + 1}{\sqrt{1 + q^2_j - \dot s^2_j }} 
- 1 \right) \right] 
\frac{ 1 +  q^2_j }
{\left[ 1 + q^2_j - \dot s ^2_j \right]^{3/2}} .
\label {HI L}
   \end{split}
\end{equation}
We do these integrals in parallel
and store their values in a lookup table.
We then use the lookup table 
and the Monte Carlo method
to estimate the mean value
\begin{equation}
\la H \ra_\beta ={}
\left\la \frac{1}{n} \sum_{k=1}^n 
\frac{C(q_{k+1}, q_k)}{  A(q_{k+1},q_k) }
\right\ra  .
\label {and measure L} 
\end{equation}
\par
\begin{figure}[H]
\begin{center}
\includegraphics[trim={0 2.75in 0 2.75in}, clip,
width=\textwidth]{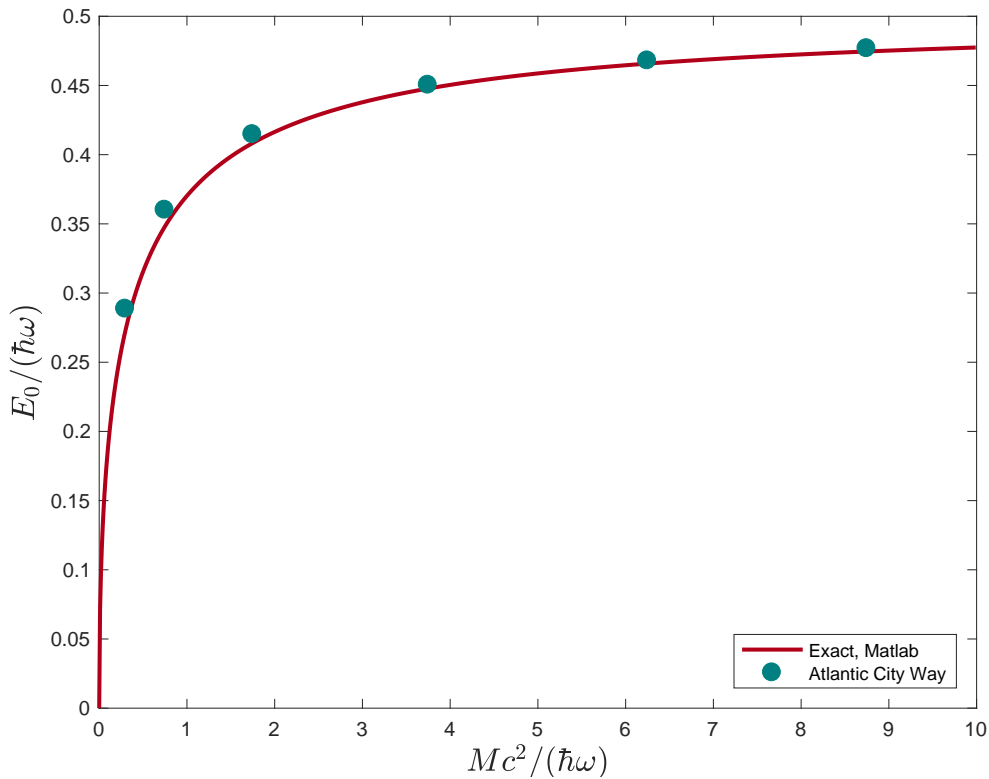}
\caption{Our Atlantic City  estimates 
(\ref{I L}--\ref{and measure L}, blue-green dots)
of the ground-state energies of the 
Born-Infeld oscillator are plotted along 
with the exact values (red curve) for 
\( 0.1 \le Mc^2/(\hbar \omega) \le 10 \),
\( a \omega = 0.1 \), and \( \beta = 10^3/M \)\@.}
\label {very awkward Born-Infeld big masses}
\end{center}
\end{figure}

We plotted our results for 
\( 0.1 \le Mc^2/(\hbar \omega) \le 10 \),
\( a \omega = 0.1 \), and \( \beta = 10^3/M \)
as green dots in 
Fig.~\ref{very awkward Born-Infeld big masses}
and listed them in 
Table~\ref{very awkward bigMasses}\@.
For comparable amounts 
of computation,
these results are not quite
as accurate as those of
Table\,\ref{bigMasses}\@.
The reason is that the 
argument of the cosine
in the formulas
(\ref {I L} \& \ref {HI L})
for \( A(q_{j+1}, q_j) \)
and \( C(q_{j+1}, q_j) \)
diverges as 
\( \dot s^2 \to q_j^2 + 1 \)\@.
The integrals converge,
but one needs more points
at small \( M \) for fixed \( a \)
and \( \omega \)\@.

\par
\begin{table}[H]
\caption{Exact (Matlab) and 
Atlantic City  results 
(\ref {I L}--\ref{and measure L})
for the ground-state energies \( E_0/(\hbar \omega) \) 
of the Born-Infeld oscillator
(\ref{Born-Infeld hamiltonian})
for \( 0.3 \le Mc^2/(\hbar \omega) \le 8.75 \)\@.}
\begin{center}
\begin{tabular}{|c|c|c|}
\hline
\( M/(\hbar \omega) \)& \( E_0/(\hbar \omega) \) exact 
  & \(E_0/(\hbar \omega) \) Atlantic City   \\
\hline
0.3&0.2731&0.2884\\
\hline
0.75&0.3482&0.3599\\
\hline
1.75&0.4084&0.4145\\
\hline
3.75&0.4478&0.4502\\
\hline
6.25&0.4658&0.4678\\
\hline
8.75&0.4746&0.4766\\
\hline
\end{tabular}
\end{center}
\label {very awkward bigMasses}
\end{table}

\section{Transition to field theory
\label{Transition to field theory}}

In this section,
we sketch how the Atlantic City
method will work in field theory.
Suppose the action is awkward,
but not very, so that 
we have a hamiltonian
\begin{equation}
H ={} H(\pi^2, (\nabla \phi)^2, \phi^2) .
\label {we have a hamiltonian}
\end{equation}
The form \( \nabla \phi^2 \equiv (\nabla \phi)^2 \)
follows from rotational invariance.
The path integral 
for the partition function is
\begin{equation}
   \begin{split}
Z(\beta)
={}& \int \exp \lt\{ \int^\beta \lt[ i \dot \phi \pi  {} 
- H(\pi^2, \nabla \phi^2, \phi^2) 
\rt] d^4x \rt\} D\phi D\pi .
\label {for the partition function is}
   \end{split}
\end{equation}
\par
We derive this path integral
from integrals of products
of matrix elements like
\begin{equation}
\la \phi (t+a) | \pi \ra 
\la \pi | e^{-aH} | \phi(t) \ra ,
\label {from matrix elements like}
\end{equation}
and approximate it
on a \( n^4 \) lattice with spacing \( a \) 
and \( \beta = n a \) as
\begin{equation}
   \begin{split}
Z(\beta)
\approx {}& \prod_{i,j,k,\ell=1}^n
\int d \phi_{i,j,k,\ell} d\pi_{i,j,k,\ell}
\exp \lt[ a^3 i (\phi_{i,j,k,\ell+1} - \phi_{i,j,k,\ell})
\pi_{i,j,k,\ell} \right. 
\\
{}& \left. 
- a^4 H(\pi^2_{i,j,k,\ell}, \nabla \phi^2_{i,j,k,\ell}, \phi^2_{i,j,k,\ell}) \rt]   
\label {with beta = n a}
   \end{split}
\end{equation}
in which the squared gradients are
\begin{equation}
   \begin{split}
(\nabla \phi)^2_{i,j,k,\ell} = {}&  
\frac{( \phi_{i+1,j,k,\ell} - \phi_{i,j,k,\ell} )^2}{a^2}
+ \frac{( \phi_{i,j+1,k,\ell} - \phi_{i,j,k,\ell} )^2}{a^2}
+ \frac{( \phi_{i,j,k+1,\ell} - \phi_{i,j,k,\ell} )^2}{a^2} .
\label {in which the squared gradients are}
   \end{split}
\end{equation}
The lookup tables
are three dimensional with entries
\begin{equation}
A(\phi_+, \phi, \nabla \phi^2)
={} \int d\pi \, \cos[a^3(\phi_+ - \phi) \pi]
e^{- a^4 H(\pi^2, \nabla \phi^2, \phi^2)} 
\label {is three dimensional with entries}
\end{equation}
and if one seeks to compute the mean value
of the energy density
\begin{equation}
C(\phi_+, \phi, \nabla \phi^2)
={} \int d\pi \, H(\pi^2, \nabla \phi^2, \phi^2)
\cos[a^3(\phi_+ - \phi) \pi]
e^{- a^4 H(\pi^2, \nabla \phi^2, \phi^2)} .
\label {test these formulas}
\end{equation}
\par
\begin{figure}[h!]
\begin{center}
\includegraphics[trim={0 2.65in 0 2.75in}, clip, 
width=\textwidth]{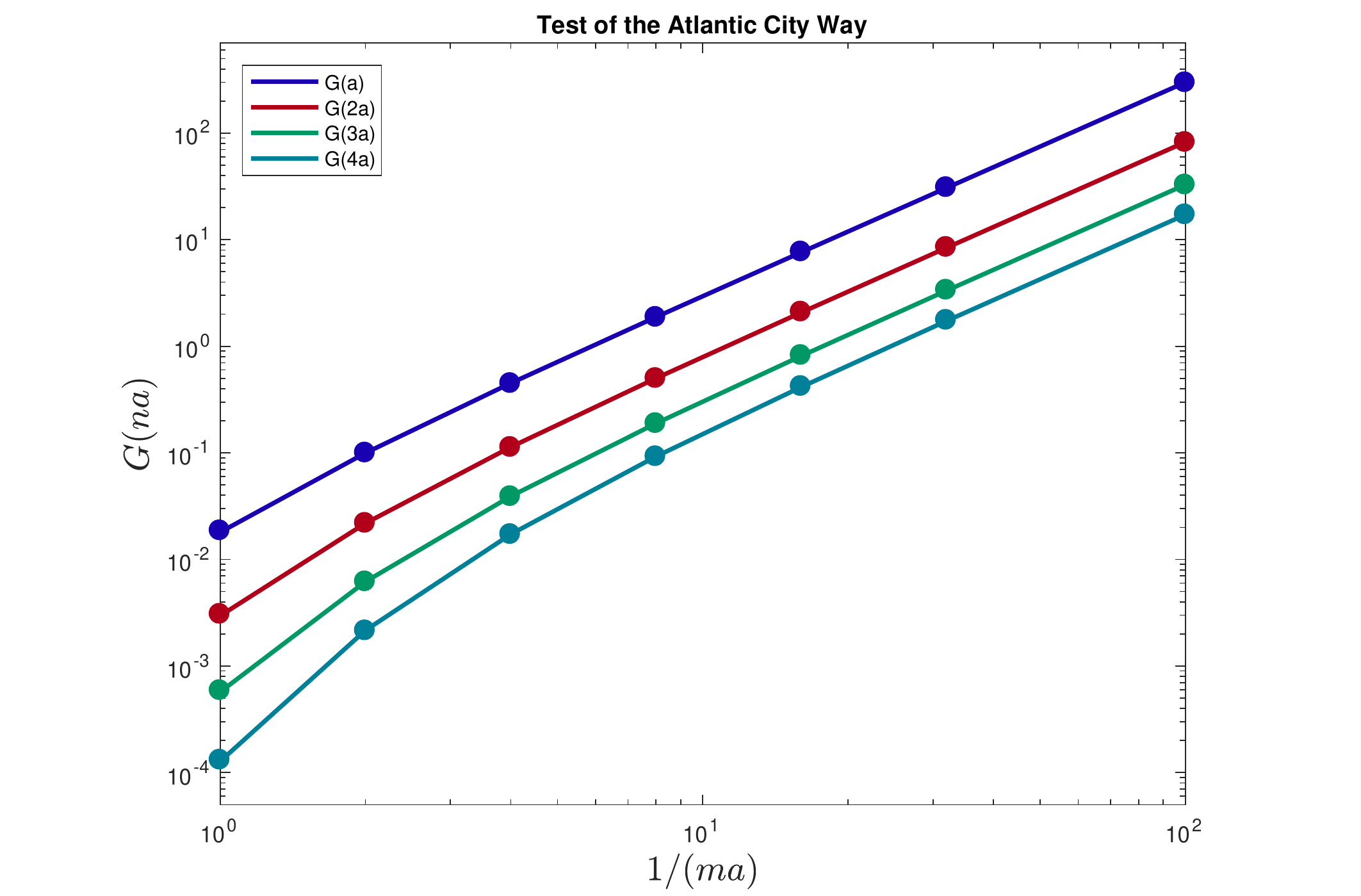}
\caption{The exact infinite-lattice euclidian Green's functions
\( G(na,0) \) (solid) for \( n = 1 \) (blue), 
2 (red), 3 (green), and 4 (blue green)
and our Atlantic City estimates  
done on an \(80^4\) lattice (dots).}
\label {AC graph}
\end{center}
\end{figure}

We have tested the Atlantic City way
by using it 
to estimate the euclidian Green's functions
\begin{equation}
G(x,y) ={} \la 0 | \mathcal{T} \lt[
\phi(x) \phi(y) \rt] | 0 \ra 
\label {two-point Green's function}
\end{equation}
of the free field theory (\ref {free L})\@.
Using parallel computing, we 
made three-dimensional lookup tables
of the values of \( A(\phi_+, \phi, \nabla \phi^2) \)
for \( m = 1 \) and 
lattice spacings \( a = 1, 1/2, 1/4, ..., 1/32, \) and 0.01\@.
We then used the standard Monte Carlo method
to estimate the path integrals
\begin{equation}
G(x,y) ={} \frac{\int \phi(x) \, \phi(y) 
\, A(\phi_+, \phi, \nabla \phi^2) \, D\phi}
{\int A(\phi_+, \phi, \nabla \phi^2) \, D\phi}
\label {standard Monte Carlo method}
\end{equation}
on lattices as big as \( 80^4 \)\@.
Our Atlantic City estimates are listed
in Table~\ref{AC lattice}
and plotted in Fig.~\ref{AC graph}\@.

\begin{table}[h!]
\caption{Atlantic City way estimate
of the free-field Green's function for \( m = 1 \)}
\begin{center}
\begin{tabular}{|c|c|c|c|c|}
\hline
\(a\) & \(G(a)\) & \(G(2a)\) & \(G(3a)\) & \(G(4a)\) \\
\hline
1.0 & 0.01853  & 0.003051 &  0.0005872 & 0.0001307 \\
\hline
0.5 & 0.09947 & 0.02177 &  0.006156 & 0.002135  \\
\hline
0.25 & 0.4462 & 0.1122 & 0.03877 & 0.01709 \\
\hline
0.125 & 1.8643 & 0.4976 & 0.1878 & 0.0919 \\
\hline
0.0625 & 7.6632 & 2.0909 & 0.8185 & 0.4185  \\
\hline
0.03125 & 30.7463 & 8.4501 & 3.3537 & 1.7486 \\
\hline
0.01 & 296.3381 & 81.6334 & 32.5551 & 17.0888 \\
\hline
\end{tabular}
\end{center}
\label {AC lattice}
\end{table}

\par
On an infinite lattice of spacing \( a \),
the exact euclidian Green's function
for \( y \) at the origin and \( x = (na, 0, 0, 0) \) 
is~\cite{Rothe3}
\begin{equation}
G(na) ={} G_{\mbox{{\scriptsize lat}}}(na) ={} \frac{1}{a^2}
\int_{-\pi}^\pi
\frac{e^{i p_1 n}}{\lt[ a^2m^2 
+ \sum_i 4 \sin^2(p_i/2) \rt]}
\frac{d^4p}{(2\pi)^4} .
\label{infinite lattice of spacing}
\end{equation}
We used Mathematica to numerically integrate
this expression and got
the values listed in Table~\ref{Exact lattice}\@.
Fig.~\ref{AC graph} shows that 
the agreement with our Atlantic City
estimates is excellent.

\begin{table}[h!]
\caption{Exact infinite-lattice free-field Green's function for \( m = 1 \):}
\begin{center}
\begin{tabular}{|c|c|c|c|c|}
\hline
\(a\) & \(G(a)\) & \(G(2a)\) & \(G(3a)\) & \(G(4a)\) \\
\hline
1.0 & 0.0180008  & 0.00296172 &  0.000571368 & 0.000127571 \\
\hline
0.5 & 0.0997255 & 0.0218474 &  0.00618522 & 0.00215997  \\
\hline
0.25 & 0.450334 & 0.113275 & 0.0391267 & 0.0172243 \\
\hline
0.125 & 1.87841 & 0.500742 & 0.189039 & 0.0924796 \\
\hline
0.0625 & 7.61685 & 2.07525 & 0.811638 & 0.414713  \\
\hline
0.03125 & 30.59684 & 8.399072 & 3.327155 & 1.728039 \\
\hline
0.01 & 299.265 & 82.3963 & 32.8142 & 17.1637 \\
\hline
\end{tabular}
\end{center}
\label {Exact lattice}
\end{table}

\section{Summary
\label{Summary}}

We divide the actions 
of theories of scalar fields
into three classes---graceful, 
awkward, and very awkward.
An action is graceful if it is quadratic
in the time derivatives of the fields,
which then are 
linearly related to the momenta,
the fields, and
their spatial derivatives.
Its partition function is
a path integral over the fields
with a positive weight function.
An action is awkward if it
is not quadratic
in the time derivatives of the fields but is simple
enough for one to find its hamiltonian.
One typically can't integrate over the momenta,
and the partition function is a path integral
over the fields and their momenta
with a complex weight function.
An action is very awkward action if 
the equations for the time derivatives 
are worse than quartic,
and one can't find its hamiltonian.
\par
We have shown how to write 
the partition function as a
euclidian path integral 
when one doesn't know
the hamiltonian.
We also have shown how
to estimate euclidian path integrals
that have weight functions 
that assume negative or complex values.
One integrates numerically 
over the momenta if the action
is awkward or over
auxiliary time derivatives
if it is very awkward.
The numerical integrations
are ideally suited to parallel computation.
One stores the values of the integrals
in lookup tables
and uses them to guide standard 
Monte Carlos. 
We demonstrated and tested
this Atlantic City method
on the Born-Infeld oscillator
by treating its action both
as awkward and as very awkward.
We sketched how to extend
this method to field theory
and tested it by computing
the known euclidian Green's functions
of the free field theory.
\par
Theories with graceful actions
have infinite energy densities. 
The Atlantic City method lets one
estimate the energy density
of theories with awkward
or very awkward 
actions, some of which 
may have finite or less than quartically divergent
energy densities\,\cite{Bender1990, *Cahill2013NA, 
*PhysRevD.88.125014NA}\@.
The Atlantic City method also 
provides a way to estimate
the acceleration of the scale factor
\( a(t) \) which in terms of the 
energy-momentum tensor \( T_{ij} \)
and its trace is
\begin{equation}
\frac{\ddot a(t)}{a(t)} ={}
- \frac{8 \pi G}{3} \lt (T_{00} + \frac{T}{2} \rt)  
\label {ddot a}
\end{equation}
in theories with awkward
or very awkward 
actions.
So the Atlantic City way of estimating
path integrals may lead
to a theory of dark energy. 
\par 
The approximation of multiple
integrals with weight functions
that assume negative or complex values
is a long-standing problem 
in applied mathematics,
called the sign problem.
The Atlantic City method
solves it for problems in which
numerical integration leads to 
a positive weight function.
\par
In the course of this paper,
we incidentally showed that
the classical hamiltonian
of the Nambu-Got{\={o}} string
vanishes identically and that
the folk theorem linking 
path integrals in real and imaginary
time can fail when the action
is awkward or very awkward.

\begin{acknowledgments}
We are grateful to Edward Witten
for shrinking our derivation 
of the vanishing of the Nambu-Got{\={o}}
energy to a single equation 
(\ref {NG energy vanishes})\@.
Conversations with Michael Creutz, 
Akimasa Miyake, 
Sudhakar Prasad, Shashank Shalgar,
and Daniel Topa also
advanced this work.  
We did some of the
numerical computations at 
the Korea Institute for Advanced Study
and most of them
at the  National Energy Research Scientific Computing Center, a DOE Office of Science User Facility supported by the Office of Science of the U.S. Department of Energy under Contract 
No. DE-AC02-05CH11231.
\end{acknowledgments}

\appendix*

\section{Ratios of complex Monte Carlos are unreliable
\label {Ratios of complex Monte Carlos are unreliable}}

The mean value of an observable
\( A[\phi] \) at inverse temperature \( \b \) is
\begin{equation}
   \begin{split}    
\la A[\phi] \ra ={}& \frac{ \tr \, A[\phi] e^{- \b H} }
{ \tr \, e^{- \b H} } \\
={}&
\int A[\phi] \exp \lt[ \int    \lt( i \, \dot \phi_j 
-  \dot \psi_j \rt) \frac{ \p  L }{ \p \dot \psi_j }  
+ L(\phi, \dot \psi) \, d^4x \rt] 
{\lt| \det \lt( \frac{ \p^2  L }
 { \p \dot \psi_k \p \dot \psi_\ell }  \rt) \rt| }
D\phi D\dot\psi 
\\
{}& \lt/ \int \exp \lt[ \int  \lt( i \, \dot \phi_j 
-  \dot \psi_j \rt) \frac{ \p  L }{ \p \dot \psi_j }  
+ L(\phi, \dot \psi) \, d^4x \rt] 
{\lt| \det \lt( \frac{ \p^2  L }
 { \p \dot \psi_k \p \dot \psi_\ell }  \rt) \rt| } 
 D\phi D\dot\psi \rt. .
 \label {the mean value}
\end{split}
\end{equation} 
The complex action 
\begin{equation}
S = {} \int \lt( i \, \dot \phi_j 
	-  \dot \psi_j \rt) \frac{ \p  L }{ \p \dot \psi_j }  
	+ L(\phi, \dot \psi)\, d^4x
	\label {the complex action}
\end{equation}
oscillates and does not give us a probability distribution
unless we can integrate over $D\dot\psi$\@.
One can write the mean value (\ref {the mean value})
as a ratio of mean values
\begin{equation}
\begin{split}    
\la A[\phi] \ra ={}& \lt\la A[\phi]
\exp \lt[ \int  i \, \dot \phi_j \,
 \frac{ \p  L }{ \p \dot \psi_j }  \, d^4x \rt]
\rt\ra
 \lt / 
\lt\la 
\exp \lt[ \int   i \, \dot \phi_j 
\, \frac{ \p  L }{ \p \dot \psi_j } \, d^4x \rt]
\rt\ra \rt. \\
={}& \int A[\phi]
\exp \lt[ \int  i \, \dot \phi_j \,
 \frac{ \p  L }{ \p \dot \psi_j }  \, d^4x \rt]
P(\phi, \dot \psi) D\phi D\dot\psi \\
{}& \quad \left/ \int \exp \lt[ \int   i \, \dot \phi_j 
\, \frac{ \p  L }{ \p \dot \psi_j } \, d^4x \rt]
P(\phi, \dot \psi) \, D\phi D\dot\psi \right.
\label {observable}
\end{split}
\end{equation}
in which the functional \( P(\phi, \dot \psi) \)
is a normalized probability distribution
\begin{equation}
   \begin{split}
   P(\phi, \dot \psi) ={}&
\exp \lt[ \int  \lt( L(\phi, \dot \psi) 
-  \dot \psi_j  \, \frac{ \p  L }{ \p \dot \psi_j }  \rt)
 d^4x \rt] 
\lt| \det \lt( \frac{ \p^2  L }
 { \p \dot \psi_k \p \dot \psi_\ell }  \rt) \rt|
\\
{}& \lt / \int
 \exp \lt[ \int   \lt( L(\phi, \dot \psi) 
 -  \dot \psi_j \, \frac{ \p  L }{ \p \dot \psi_j }  \rt)
  d^4x \rt] 
\lt| \det \lt( \frac{ \p^2  L }
 { \p \dot \psi_k \p \dot \psi_\ell }  \rt) \rt|
 \, D\phi D\dot\psi  \rt. .
 \label {first probability distribution}
   \end{split}
\end{equation}
\par
Although in principle one can use
the Monte Carlo method~\cite{CahillXIV}
to estimate
the numerator \( N \) and the denominator \( D \)
of the ratio (\ref {observable}),
both \( N \) and  \( D \) are
the mean values of complex oscillating
functionals.
In many cases 
of interest, both \( N \) and \( D \)
are smaller than the measurement errors 
\( \delta N \) and \( \delta D \) 
in computations of reasonable lengths.
The error in the observable \( A[\phi] \) is
\begin{equation}
\delta \la A[\phi] \ra ={}
\delta \frac{N}{D} = 
\frac{ \delta N }{D} - 
\frac{ N }{ D^2 } \delta D =
\frac{ \delta N }{D} -  \la A[\phi] \ra
\frac{ \delta D }{ D } ,
\label {error in N/D}
\end{equation}
and both \( N \) and \( D \) often are zero
in the limit in which \( \beta \to \infty \)\@.
\par
For instance, suppose we apply the technique
(\ref {observable}--\ref {first probability distribution})
to the computation of the ground-state energy
\( \la H \ra = N/D \)
of the harmonic oscillator
in which the numerator is
\begin{align} 
N = {}& \int \left(  \frac{p^2}{2m}
   + \half m \omega^2 q^2 \right)
   \exp\left[\int \left( i \dot q p - \frac{p^2}{2m}
   - \half m \omega^2 q^2 \right) dt \right] Dp Dq \nn\\
  {}& \left/  \int \exp \left[ \int  \left( - \frac{p^2}{2m}
   - \half m \omega^2 q^2 \right) dt \right]  Dp Dq \right. ,
   \label {computation of the numerator}
\end{align}
the denominator \( D\) is
\begin{align} 
D = {}& \int \exp\left[\int \left( i \dot q p - \frac{p^2}{2m}
   - \half m \omega^2 q^2 \right) dt \right] Dp Dq \nn\\
  {}& \left/  \int \exp \left[ \int  \left( - \frac{p^2}{2m}
   - \half m \omega^2 q^2 \right) dt \right]  Dp Dq \right. ,
   \label {computation of the denominator}
\end{align}
and the measure \( Dp Dq \) is
\begin{equation}
Dp Dq = \prod_{j=1}^n \frac{1}{2\pi} dp_j dq_j .
\label {Dp Dq}
\end{equation}
In the continuum limit 
($n\rightarrow\infty$, $dt\rightarrow 0$, with $\beta = n\:dt$ fixed),
the numerator \( \mathcal{N} \)
of the denominator \( D \equiv \mathcal{N} / \mathcal{D} \) 
of the ratio \( N/D \)
is the partition function 
\begin{align}
\mathcal{N} = {}& Z(\beta) = \tr \, e^{-\beta H} \nn\\
= {}& \int \exp\left[ \int \left( i \dot q p - \frac{p^2}{2m}
   - \half m \omega^2 q^2 \right) dt \right] Dp Dq \nn\\
= {}& \frac{1}{2 \sinh (\beta \omega/2)} ,
   \label {The numerator of this expression is}
\end{align}
and the denominator is
\begin{align} 
\mathcal{D} ={}& 
\int \exp \left[ \int  \left( - \frac{p^2}{2m}
   - \half m \omega^2 q^2 \right) dt \right]  
   Dp Dq \nn  \\
   ={}& \prod_{j=1}^n \int  \frac{dp_j dq_j}{2\pi}
   \exp \left[ \sum_{j=1}^n \left( - \frac{p^2_j}{2m}
   - \half m \omega^2 q^2_j \right) \frac{\beta}{n}
    \right] \nn  \\
   ={}& \left(\frac{1}{2\pi}\right)^n
   \left(\frac{2\pi m n}{\beta} \right)^{n/2}
 \prod_{j=1}^n \int dq_j 
   \exp \left[ \sum_{j=1}^n  \left( 
      - \half m \omega^2 q^2_j \right) \frac{\beta}{n}
       \right] \nn  \\  
 ={}& \left(\frac{1}{2\pi}\right)^n
 \left(\frac{2\pi m n}{\beta} \right)^{n/2}   
\left(\frac{2\pi n}{m \omega^2 \beta} \right)^{n/2}    
= \left( \frac{n}{\beta \omega} \right)^n  .
  \end{align}
This denominator goes to infinity as \( n \to \infty \) and 
\( \beta/n \to 0 \) for any \( \beta \ne 0 \)\@.
So the denominator \( D \) vanishes
\begin{equation}
D =  \frac{1}{2 \sinh (\beta \omega/2) \, 
\mathcal{D}} = 0 .
\label {So the denominator =0}
\end{equation}
The numerator also vanishes,
so the ratio \( \la H \ra = N/D \)
is hard to estimate, being \( 0 /0 \)\@.
\par
Thus the double-ratio trick
(\ref {observable}--\ref {first probability distribution})
is not in general reliable.

\bibliography{physics}

\begin{thebibliography}{19}%
\makeatletter
\providecommand \@ifxundefined [1]{%
 \@ifx{#1\undefined}
}%
\providecommand \@ifnum [1]{%
 \ifnum #1\expandafter \@firstoftwo
 \else \expandafter \@secondoftwo
 \fi
}%
\providecommand \@ifx [1]{%
 \ifx #1\expandafter \@firstoftwo
 \else \expandafter \@secondoftwo
 \fi
}%
\providecommand \natexlab [1]{#1}%
\providecommand \enquote  [1]{``#1''}%
\providecommand \bibnamefont  [1]{#1}%
\providecommand \bibfnamefont [1]{#1}%
\providecommand \citenamefont [1]{#1}%
\providecommand \href@noop [0]{\@secondoftwo}%
\providecommand \href [0]{\begingroup \@sanitize@url \@href}%
\providecommand \@href[1]{\@@startlink{#1}\@@href}%
\providecommand \@@href[1]{\endgroup#1\@@endlink}%
\providecommand \@sanitize@url [0]{\catcode `\\12\catcode `\$12\catcode
  `\&12\catcode `\#12\catcode `\^12\catcode `\_12\catcode `\%12\relax}%
\providecommand \@@startlink[1]{}%
\providecommand \@@endlink[0]{}%
\providecommand \url  [0]{\begingroup\@sanitize@url \@url }%
\providecommand \@url [1]{\endgroup\@href {#1}{\urlprefix }}%
\providecommand \urlprefix  [0]{URL }%
\providecommand \Eprint [0]{\href }%
\providecommand \doibase [0]{http://dx.doi.org/}%
\providecommand \selectlanguage [0]{\@gobble}%
\providecommand \bibinfo  [0]{\@secondoftwo}%
\providecommand \bibfield  [0]{\@secondoftwo}%
\providecommand \translation [1]{[#1]}%
\providecommand \BibitemOpen [0]{}%
\providecommand \bibitemStop [0]{}%
\providecommand \bibitemNoStop [0]{.\EOS\space}%
\providecommand \EOS [0]{\spacefactor3000\relax}%
\providecommand \BibitemShut  [1]{\csname bibitem#1\endcsname}%
\let\auto@bib@innerbib\@empty
\bibitem [{\citenamefont {Ade}\ and\ \citenamefont {others
  (Planck~Collaboration)}(2013)}]{PlanckCosmologicalNA}%
  \BibitemOpen
  \bibfield  {author} {\bibinfo {author} {\bibfnamefont {P.}~\bibnamefont
  {Ade}}\ and\ \bibinfo {author} {\bibnamefont {others
  (Planck~Collaboration)}},\ }\href@noop {} {\bibfield  {journal} {\bibinfo
  {journal} {Astron.\ Astrophys.\ {\bf 566}, A54 (2014)}\ } (\bibinfo {year}
  {2013})}\BibitemShut {NoStop}%
\bibitem [{\citenamefont {Weinberg}(1995)}]{Weinberg1995IX}%
  \BibitemOpen
  \bibfield  {author} {\bibinfo {author} {\bibfnamefont {S.}~\bibnamefont
  {Weinberg}},\ }\href@noop {} {\emph {\bibinfo {title} {The Quantum Theory of
  Fields}}},\ Vol.~\bibinfo {volume} {I}\ (\bibinfo  {publisher} {Cambridge
  University Press},\ \bibinfo {year} {1995})\ Chap.~\bibinfo {chapter}
  {9}\BibitemShut {NoStop}%
\bibitem [{\citenamefont {Cahill}(2015)}]{Cahill1501}%
  \BibitemOpen
  \bibfield  {author} {\bibinfo {author} {\bibfnamefont {K.}~\bibnamefont
  {Cahill}},\ }\href@noop {} {\enquote {\bibinfo {title} {Path integrals for
  actions that are not quadratic in their time derivatives},}\ } (\bibinfo
  {year} {2015}),\ \Eprint {http://arxiv.org/abs/1501.00473} {arXiv:1501.00473
  [hep-th]} \BibitemShut {NoStop}%
\bibitem [{\citenamefont {Born}\ and\ \citenamefont
  {Infeld}(1934{\natexlab{a}})}]{Born:1934gh}%
  \BibitemOpen
  \bibfield  {author} {\bibinfo {author} {\bibfnamefont {M.}~\bibnamefont
  {Born}}\ and\ \bibinfo {author} {\bibfnamefont {L.}~\bibnamefont {Infeld}},\
  }\href {\doibase 10.1098/rspa.1934.0059} {\bibfield  {journal} {\bibinfo
  {journal} {Proc. Roy. Soc. Lond.}\ }\textbf {\bibinfo {volume} {A144}},\
  \bibinfo {pages} {425} (\bibinfo {year} {1934}{\natexlab{a}})}\BibitemShut
  {NoStop}%
\bibitem [{\citenamefont {Born}\ and\ \citenamefont
  {Infeld}(1934{\natexlab{b}})}]{Born:1934dia}%
  \BibitemOpen
  \bibfield  {author} {\bibinfo {author} {\bibfnamefont {M.}~\bibnamefont
  {Born}}\ and\ \bibinfo {author} {\bibfnamefont {L.}~\bibnamefont {Infeld}},\
  }\href {\doibase 10.1098/rspa.1934.0234} {\bibfield  {journal} {\bibinfo
  {journal} {Proc. Roy. Soc. Lond.}\ }\textbf {\bibinfo {volume} {A147}},\
  \bibinfo {pages} {522} (\bibinfo {year} {1934}{\natexlab{b}})}\BibitemShut
  {NoStop}%
\bibitem [{\citenamefont {Born}\ and\ \citenamefont
  {Infeld}(1935)}]{Born:1935ap}%
  \BibitemOpen
  \bibfield  {author} {\bibinfo {author} {\bibfnamefont {M.}~\bibnamefont
  {Born}}\ and\ \bibinfo {author} {\bibfnamefont {L.}~\bibnamefont {Infeld}},\
  }\href {\doibase 10.1098/rspa.1935.0093} {\bibfield  {journal} {\bibinfo
  {journal} {Proc. Roy. Soc. Lond.}\ }\textbf {\bibinfo {volume} {A150}},\
  \bibinfo {pages} {141} (\bibinfo {year} {1935})}\BibitemShut {NoStop}%
\bibitem [{\citenamefont {Bender}\ and\ \citenamefont
  {Mannheim}(2008{\natexlab{a}})}]{BenderMannheim2007}%
  \BibitemOpen
  \bibfield  {author} {\bibinfo {author} {\bibfnamefont {C.~M.}\ \bibnamefont
  {Bender}}\ and\ \bibinfo {author} {\bibfnamefont {P.~D.}\ \bibnamefont
  {Mannheim}},\ }\href@noop {} {\bibfield  {journal} {\bibinfo  {journal}
  {Phys.\ Rev.\ Letters}\ }\textbf {\bibinfo {volume} {100}},\ \bibinfo {pages}
  {110402} (\bibinfo {year} {2008}{\natexlab{a}})}\BibitemShut {NoStop}%
\bibitem [{\citenamefont {Bender}\ and\ \citenamefont
  {Mannheim}(2008{\natexlab{b}})}]{BenderMannheim2008}%
  \BibitemOpen
  \bibfield  {author} {\bibinfo {author} {\bibfnamefont {C.~M.}\ \bibnamefont
  {Bender}}\ and\ \bibinfo {author} {\bibfnamefont {P.~D.}\ \bibnamefont
  {Mannheim}},\ }\href@noop {} {\bibfield  {journal} {\bibinfo  {journal} {J.\
  Phys.\ A}\ }\textbf {\bibinfo {volume} {41}},\ \bibinfo {pages} {304018}
  (\bibinfo {year} {2008}{\natexlab{b}})}\BibitemShut {NoStop}%
\bibitem [{\citenamefont {Dirac}(1950)}]{Dirac1950}%
  \BibitemOpen
  \bibfield  {author} {\bibinfo {author} {\bibfnamefont {P.~A.~M.}\
  \bibnamefont {Dirac}},\ }\href@noop {} {\bibfield  {journal} {\bibinfo
  {journal} {Can.\ J.\ Math.}\ }\textbf {\bibinfo {volume} {2}},\ \bibinfo
  {pages} {129} (\bibinfo {year} {1950})}\BibitemShut {NoStop}%
\bibitem [{\citenamefont {Dirac}(1958)}]{Dirac1958}%
  \BibitemOpen
  \bibfield  {author} {\bibinfo {author} {\bibfnamefont {P.~A.~M.}\
  \bibnamefont {Dirac}},\ }\href@noop {} {\bibfield  {journal} {\bibinfo
  {journal} {Proc.\ Roy.\ Soc.\ London, ser.\ A}\ }\textbf {\bibinfo {volume}
  {246}},\ \bibinfo {pages} {326} (\bibinfo {year} {1958})}\BibitemShut
  {NoStop}%
\bibitem [{\citenamefont {Dirac}(1964)}]{Dirac1964}%
  \BibitemOpen
  \bibfield  {author} {\bibinfo {author} {\bibfnamefont {P.~A.~M.}\
  \bibnamefont {Dirac}},\ }\href@noop {} {\emph {\bibinfo {title} {Lectures on
  Quantum Mechanics}}}\ (\bibinfo  {publisher} {Yeshiva University and Dover
  Publications},\ \bibinfo {year} {1964})\BibitemShut {NoStop}%
\bibitem [{\citenamefont {Cahill}(2013{\natexlab{a}})}]{CahillXVI}%
  \BibitemOpen
  \bibfield  {author} {\bibinfo {author} {\bibfnamefont {K.}~\bibnamefont
  {Cahill}},\ }\enquote {\bibinfo {title} {\textit{Physical Mathematics}},}\ \
  (\bibinfo  {publisher} {Cambridge University Press},\ \bibinfo {year}
  {2013})\ Chap.~\bibinfo {chapter} {16}\BibitemShut {NoStop}%
\bibitem [{\citenamefont {Cahill}({\natexlab{a}})}]{CahillXIXi}%
  \BibitemOpen
  \bibfield  {author} {\bibinfo {author} {\bibfnamefont {K.}~\bibnamefont
  {Cahill}},\ }\href@noop {} {\emph {\bibinfo {title} {\textit{ibid}}}},\
  Chap.~\bibinfo {chapter} {19}\BibitemShut {NoStop}%
\bibitem [{\citenamefont {Cahill}({\natexlab{b}})}]{CahillXIVi}%
  \BibitemOpen
  \bibfield  {author} {\bibinfo {author} {\bibfnamefont {K.}~\bibnamefont
  {Cahill}},\ }\href@noop {} {\emph {\bibinfo {title} {\textit{ibid}}}},\
  Chap.~\bibinfo {chapter} {14}\BibitemShut {NoStop}%
\bibitem [{\citenamefont {Rothe}(1997)}]{Rothe3}%
  \BibitemOpen
  \bibfield  {author} {\bibinfo {author} {\bibfnamefont {H.~J.}\ \bibnamefont
  {Rothe}},\ }\href@noop {} {\emph {\bibinfo {title} {Lattice Gauge Theories:
  An Introduction}}},\ \bibinfo {edition} {2nd}\ ed.\ (\bibinfo  {publisher}
  {World Scientific},\ \bibinfo {year} {1997})\ Chap.~\bibinfo {chapter}
  {3}\BibitemShut {NoStop}%
\bibitem [{\citenamefont {Boettcher}\ and\ \citenamefont
  {Bender}(1990)}]{Bender1990}%
  \BibitemOpen
  \bibfield  {author} {\bibinfo {author} {\bibfnamefont {S.}~\bibnamefont
  {Boettcher}}\ and\ \bibinfo {author} {\bibfnamefont {C.}~\bibnamefont
  {Bender}},\ }\href@noop {} {\bibfield  {journal} {\bibinfo  {journal} {J.
  Math. Phys.}\ }\textbf {\bibinfo {volume} {31(11)}},\ \bibinfo {pages} {2579}
  (\bibinfo {year} {1990})}\BibitemShut {NoStop}%
\bibitem [{\citenamefont {Cahill}(2013{\natexlab{b}})}]{Cahill2013NA}%
  \BibitemOpen
  \bibfield  {author} {\bibinfo {author} {\bibfnamefont {K.}~\bibnamefont
  {Cahill}},\ }\href@noop {} {\bibfield  {journal} {\bibinfo  {journal} {Phys.
  Rev. D}\ }\textbf {\bibinfo {volume} {87}},\ \bibinfo {pages} {065024}
  (\bibinfo {year} {2013}{\natexlab{b}})}\BibitemShut {NoStop}%
\bibitem [{\citenamefont {Cahill}(2013{\natexlab{c}})}]{PhysRevD.88.125014NA}%
  \BibitemOpen
  \bibfield  {author} {\bibinfo {author} {\bibfnamefont {K.}~\bibnamefont
  {Cahill}},\ }\href {\doibase 10.1103/PhysRevD.88.125014} {\bibfield
  {journal} {\bibinfo  {journal} {Phys. Rev. D}\ }\textbf {\bibinfo {volume}
  {88}},\ \bibinfo {pages} {125014} (\bibinfo {year}
  {2013}{\natexlab{c}})}\BibitemShut {NoStop}%
\bibitem [{\citenamefont {Cahill}(2013{\natexlab{d}})}]{CahillXIV}%
  \BibitemOpen
  \bibfield  {author} {\bibinfo {author} {\bibfnamefont {K.}~\bibnamefont
  {Cahill}},\ }\enquote {\bibinfo {title} {\textit{Physical Mathematics}},}\ \
  (\bibinfo  {publisher} {Cambridge University Press},\ \bibinfo {year}
  {2013})\ Chap.~\bibinfo {chapter} {14}\BibitemShut {NoStop}%
\end{thebibliography}%

\end{document}